\documentclass[aps,prd,twocolumn,superscriptaddress,nofootinbib,preprintnumbers]{revtex4-1}
\usepackage{graphics} 
\usepackage{epsfig}
\usepackage[usenames]{color}
\usepackage{verbatim}
\usepackage{amsmath,amsfonts,amssymb}
\usepackage{slashed}

\begin{document}

\title{Leptogenesis Via Neutrino Production During Higgs Relaxation}

\author{Lauren Pearce}
\affiliation{William I. Fine Theoretical Physics Institute, School of Physics and Astronomy, University of Minnesota, Minneapolis, MN 55455 USA}

\author{Louis Yang}
\affiliation{Department of Physics and Astronomy, University of California, Los Angeles, CA 90095-1547, USA}

\author{Alexander Kusenko}
\affiliation{Department of Physics and Astronomy, University of California, Los Angeles, CA 90095-1547, USA}
\affiliation{Kavli Institute for the Physics and Mathematics of the Universe (WPI), University of Tokyo, Kashiwa, Chiba 277-8568, Japan}

\author{Marco Peloso}
\affiliation{School of Physics and Astronomy, University of Minnesota, Minneapolis, MN 55455 USA}

\preprint{FTPI-MINN-15/21}
\preprint{UMN-TH-3434/15}

\begin{abstract}
During inflation, scalar fields, including the Higgs boson, may acquire a nonzero vacuum expectation value, which must later relax to the equilibrium value during reheating.  In the presence of the time-dependent condensate, the vacuum state can evolve into a state with a nonzero particle number.  We show that, in the presence of lepton number violation in the neutrino sector, the particle production can explain the observed matter-antimatter asymmetry of the universe. 
We find that this form of leptogenesis is particularly effective when the Higgs condensate decays rapidly and at low reheat scale.  As part of the calculation, we present some exact results for the Bogoliubov transformations for Majorana fermions with a nonzero time-dependent chemical potential, in addition to a time-dependent mass.
\end{abstract}

\maketitle

\section{Introduction}

During the inflationary era, the Higgs field may develop a stochastic distribution of vacuum expectation values (VEVs) due to the flatness of its potential~\cite{Bunch:1978yq,Linde:1982uu,Starobinsky:1994bd}, or it may be trapped in a quasi-stable minimum.  In both cases, after inflation the Higgs field relaxes to its vacuum state via a coherent motion~\cite{Enqvist:2013kaa,Kusenko:2014lra,Yang:2015ida}.  At large VEVs, the Higgs field may be sensitive to physics beyond the Standard Model, and new terms in the Lagrangian, such as those considered in models of spontaneous baryogenesis, can generate an effective potential for baryon and lepton number~\cite{Cohen:1987vi,Dine:1990fj}.  These terms couple the time-dependent scalar condensate to the lepton (and baryon) number currents.  Consequently, the thermal bath of particles produced by reheating can lower its energy by converting particles into antiparticles, through scattering involving neutrinos, whose Majorana mass violates lepton number~\cite{Kusenko:2014lra,Yang:2015ida}.  This scenario can explain the observed matter-antimatter asymmetry of the universe; it requires sufficiently fast reheating, such that the plasma forms before Higgs relaxation is complete, which restricts the possible parameter space. In addition to the Higgs boson, an axion or a Majoron relaxation could generate the baryon asymmetry of the universe~\citep{Ibe:2015nfa,Kusenko:2014uta,Yang:2015ida}.

However, the relaxation of the Higgs vacuum expectation value itself results in particle production.  Generically, a classically evolving background scalar field coupled to quantum fields results in particle production; this can be understood as a consequence of the fact that the initial vacuum state (which is annihilated by the appropriate annihilation operators at $t=0$) is not annihilated by the appropriate annihilation operators at later times~\cite{Birrell:1982ix,Traschen:1990sw,Dolgov:1989us}.  (More specifically, the time-dependent background mixes positive and negative energy solutions of the field equations, and so an initially diagonal Hamiltonian is non-diagonal at later times.)  This can result in the production of both scalar bosons~\cite{Traschen:1990sw,Dolgov:1989us} and fermions~\cite{Greene:2000ew,Giudice:1999fb}, provided that the classical scalar field is coupled to both.  This has been explored extensively with respect to the inflaton (e.g.,~\cite{Kofman:1994rk,Shtanov:1994ce,Kofman:1997yn}).  In this work, we calculate the excess of neutrinos over antineutrinos produced by the evolving Higgs condensate in the presence of chemical potential, generated by higher order terms in the Lagrangian, which distinguishes particles from antiparticles.

During the oscillations of the Higgs condensate, the effective chemical potential changes sign, which alternates whether the production of neutrinos or antineutrinos is favored.  Therefore, the maximal asymmetry is produced with the Higgs condensate decays quickly, which minimizes this wash out.  Furthermore, this mechanism favors a low reheating scale, which minimizes entropy production.

The outline of this paper is as follows: In the next section, we introduce our model, including the $\mathcal{O}_6$ operator which gives rise to an effective chemical potential for lepton number.  Subsequently, we derive an effective Lagrangian by integrating out the weakly interacting right-handed neutrino states, and we specialize to the case of a single fermion family.  In section \ref{sec:Quantization}, we quantize this system and find the Bogoliubov transformation equations which describe particle production.   Following this, we define the occupation number of the physical eigenstates and lepton number.  Finally, present a numerical analysis of our model, which demonstrates that resulting asymmetry can be sufficiently large to account for the observed matter-antimatter asymmetry.

\section{The Lagrangian}
\label{sec:Lagrangian}

In this section, we introduce the model Lagrangian.  We begin with the action in general curved spacetime
\begin{equation}
S = \int d^4x \sqrt{-g} \mathcal{L}
\end{equation}
with a Lagrangian 
\begin{equation}
\mathcal{L} = \mathcal{L}_H + \mathcal{L}_{\ell} + \mathcal{L}_{\mathcal{O}_6} + \mathcal{L}_\mathrm{SM},
\end{equation}
where we use $\mathcal{L}_H$ to denote the Higgs sector contribution, $\mathcal{L}_{\ell}$ to denote the lepton sector contribution, $\mathcal{L}_{\mathrm{O}_6}$ to denote higher dimensional operators which will generate an effective chemical potential for baryon and lepton number, and $\mathcal{L}_\mathrm{SM}$ represents the Standard Model contributions that do not appear in $\mathcal{L}_H$ or $\mathcal{L}_\ell$.  We consider an expanding FLRW spacetime with signature $(+,-,-,-)$.

The purely Higgs sector contribution is 
\begin{align}
\mathcal{L}_H = g^{\mu \nu} \partial_\mu \Phi^\dagger \partial_\nu \Phi - V_\phi(\Phi,T), 
\end{align}
where $V_\phi(\Phi, T)$ is the Higgs potential, including any relevant loop and finite temperature corrections.  We note that, as with the models discussed in \cite{Kusenko:2014lra,Yang:2015ida}, the potential $V_\phi(\Phi,T)$ may require higher dimensional operators involving the Higgs field $\Phi$ (and possibly the inflaton field $I$) in order to suppress isocurvature perturbations resulting from variations in the produced baryon density~\cite{1987Natur.327..210P,Enqvist:1998pf,Harigaya:2014tla}.  The Higgs sector is discussed in more detail in subsection \ref{subsec:Higgs_Sector} below.

The lepton sector Lagrangian includes the terms
\begin{align}
 & \mathcal{L}_{\ell}=i\sum\bar{L}\left(g^{\mu\nu}\tilde{\gamma}_{\mu}\partial_{\nu}+\dfrac{3}{2}g^{00}\dfrac{a^{\prime}}{a}\tilde{\gamma}_{0}\right)L\nonumber \\
 & +i\sum\bar{N}_{R}\left(g^{\mu\nu}\tilde{\gamma}_{\mu}\partial_{\nu}+\dfrac{3}{2}g^{00}\dfrac{a^{\prime}}{a}\tilde{\gamma}_{0}\right)N_{R}-\sum y_{\ell}\bar{L}\Phi\ell_{R}\nonumber \\
 & -\sum y_{\nu}\bar{L}\Phi N_{R}-\sum\dfrac{M_{N}}{2}\overline{(N_{R})^{c}}N_{R}+h.c.,
\label{eq:lepton_sector_lagrangian}
\end{align}
where $L$ denotes left-handed lepton doublets, $\ell$ right-handed charged leptons, and $N_R$ right-handed neutrinos, and we implicitly sum over indices and families.  The gamma matrices in the FRW metric are related to those in flat space time by $\tilde{\gamma}_\mu = a \gamma_{\mu}$.  We note in Eq.~\eqref{eq:lepton_sector_lagrangian} the effect of the spin connection evaluated on the FLRW background.  This Lagrangian will be discussed further in subsection \ref{subsec:Neutrino_Sector} below.

The third part of the Lagrangian is the higher dimensional operator
\begin{align}
\mathcal{L}_{\mathcal{O}_6} &= -\dfrac{\Phi^2}{M^2} \partial_\mu j_{B+L}^\mu,
\label{eq:intro_O6}
\end{align}
where $j_L^\mu$ is the lepton current density.  One possibility for generating an operator of this form is to couple the Higgs field $\Phi$ to the $\mathrm{SU}_\mathrm{L}(2) \times \mathrm{U}_\mathrm{Y}(1)$ gauge fields $\mathsf{A}$ and $\mathsf{B}$ by
\begin{equation}
\mathcal{O}_6 = - \dfrac{\Phi^2}{M^2} \dfrac{n_g}{32 \pi^2} \left( g_2^2 \epsilon^{\mu \nu \alpha \beta} \mathsf{A}^{a}_{\mu \nu} \mathsf{A}^{a}_{ \alpha \beta} - g_1^2 \epsilon^{\mu \nu \alpha \beta} \mathsf{B}_{\mu \nu} \mathsf{B}_{\alpha \beta} \right),
\end{equation}
which can be written in the form of \label{eq:intro_O6} using the electroweak anomaly equation~\cite{Cohen:1987vi,Dine:1990fj}.  This transformation requires the electroweak sphalerons to be in thermal equilibrium, which may not be satisfied here, although the situation is complicated by the time-dependent Higgs VEV.  For these reasons, we discuss other ways of generating this operator in Appendix \ref{ap:Generating_O_6_operator}.

We will discuss the role of this term further in subsection \ref{subsec:O6_Operator}; for now, we note only that in the presence of this term, the Higgs evolution induces a chemical potential that distinguishes particles from antiparticles.

Before we discuss each component separately, we will first rewrite the action using conformal time, such that the metric is $g_{\mu \nu} = a^2(\eta) \eta_{\mu \nu}$, and\footnote{Throughout this paper, we will use primes to denote differentiation with respect to conformal time, dots to denote differentiation with respect to physical time, tildes to denote comoving quantities, and hats to denote two-component fields.  Where necessary, we will use bars in dummy variables.}   
\begin{equation}
\eta = \int_0^t \dfrac{d\bar{t}}{a(\bar{t})},
\end{equation}
It will be convenient to define the ``comoving" fields, 
\begin{align}
\tilde{\phi} &= a \phi \nonumber \\
\tilde{\psi} &= a^{3 \slash 2} \psi
\end{align}
such that we can write 
\begin{align}
S = \int d^4x \left( \tilde{\mathcal{L}}_H + \tilde{\mathcal{L}}_\ell + \tilde{\mathcal{L}}_{\mathcal{O}_6} + \tilde{\mathcal{L}}_\mathrm{SM}\right)
\end{align}
where
\begin{align}
\tilde{\mathcal{L}}_H &=  \partial^\mu \tilde{\Phi} \partial_\mu \tilde{\Phi} - \dfrac{a^{\prime \prime}}{a} \tilde{\Phi}^{ 2}  - \tilde{V}_\phi(\tilde{\Phi},T),  \nonumber \\
\tilde{\mathcal{L}}_\ell &= i\sum \bar{\tilde{L}} \slashed{\partial} \tilde{L} + i\sum \bar{\tilde{N}}_R \slashed{\partial} \tilde{N}_R -\sum  y^\prime \bar{\tilde{L}} \tilde{\Phi} \tilde{\ell}_{R}  \nonumber \\
& \qquad - \sum y \bar{\tilde{L}} \tilde{\Phi} \tilde{N}_{R} - \sum \dfrac{a M_{N}}{2} \overline{(\tilde{N}_{R})^{c}} \tilde{N}_{R} + h.c., \nonumber \\
\tilde{\mathcal{L}}_{\mathcal{O}_6} &= -  \dfrac{a^4  \tilde{\Phi}^{2}}{ M^2} \partial_\mu j_{B+L}^\mu, \nonumber \\
\tilde{\mathcal{L}}_\mathrm{SM} &= a^4 \mathcal{L}_\mathrm{SM}.
\label{eq:L_hats}
\end{align}
In the first equation, we have defined a comoving potential $\tilde{V}_\phi = a^4 V_\phi$.

In the next subsection, we consider how the Higgs field might acquire a large VEV during inflation, which relaxes to its equilibrium value during reheating.  Then we consider how this affects the quadratic terms in the lepton sector; subsequently, we demonstrate that when the Higgs VEV is in motion the $\mathcal{O}_6$ operator produces a chemical potential for baryon and lepton number.  Finally, we gather together the relevant contributions to the Lagrangian in the final subsection.

\subsection{The Higgs Sector}
\label{subsec:Higgs_Sector}

The Standard Model Higgs boson has the tree-level potential 
\begin{equation}
V_\phi(\Phi) = m^2 \Phi^\dagger \Phi + \lambda (\Phi^\dagger \Phi)^2,
\end{equation}
where $\Phi$ is the Higgs $\mathrm{SU}(2)$ doublet.  The parameters $m$ and $\lambda$, although constant at tree-level, are modified by both loop and finite temperature corrections.  For the experimentally preferred top quark mass and Higgs boson mass, loop corrections result in a negative running coupling $\lambda$ at sufficiently large vacuum expectation values (VEVs), with the result that the $\sqrt{\left< \phi^2 \right>} = v_\mathrm{EM} = 246 \; \mathrm{GeV}$ minimum is metastable at zero temperature~\cite{Degrassi:2012ry}.  We note, though, that a stable vacuum is possible within current experimental uncertainties~\cite{Degrassi:2012ry}, and the stability of the potential is also sensitive to Planck-scale corrections~\cite{Branchina:2014rva}. 

Therefore, the running quartic coupling produces a shallow potential, and consequently, the Higgs field may develop a large VEV during inflation due to quantum fluctuations~\cite{Enqvist:2013kaa}.  Qualitatively, the scalar field in a de Sitter space can develop a large VEV via quantum effects, such as Hawking-Moss instantons~\cite{Bunch:1978yq,Hawking:1981fz} or stochastic growth~\cite{Linde:1982uu,Starobinsky:1982ee,Vilenkin:1982wt}.  Subsequently, the field would relax to its equilibrium value via a classical motion on the time scale $\sim (d^2 V \slash d\phi^2)^{-1 \slash 2}$, unless Hubble friction delays this relaxation.  If $H_I \gg \sqrt{ d^2 V \slash d\phi^2}$ then quantum jumps occur frequently enough to maintain a large VEV.

Alternatively, the Higgs potential is sensitive to higher-dimensional operators at large vacuum expectation values, which can have the effect of lifting the second minimum, stabilizing the electroweak vacuum.  During inflation, the Higgs field may have a stochastic distribution of VEVs similar to that of the inflaton itself in chaotic inflation models.  During inflation, sufficiently large VEVs evolve towards the false vacuum from above, and then remain trapped in this false vacuum until destabilized by thermal corrections during reheating.  Subsequently, the field rolls to the global minimum, until electroweak symmetry is broken at a significantly later time.

Therefore, it is quite natural to consider scenarios in which the Higgs field has a large vacuum expectation value during inflation, which subsequently relaxes to its equilibrium value.  Both of the above scenarios have been explored previously~\cite{Kusenko:2014lra,Yang:2015ida}.  Here, we consider the classical motion of the Higgs field towards equilibrium generically, without specifying the mechanism which generates the initial large vacuum expectation value.

We note that if the field has expectation value 
\begin{equation}
\left< \Phi \right> = \dfrac{1}{\sqrt{2}} \begin{pmatrix} v(t) \\ 0 \end{pmatrix},
\end{equation}
then the comoving field has expectation value
\begin{equation}
\left< \tilde{\Phi} \right> = \dfrac{1}{\sqrt{2}} \begin{pmatrix} \tilde{v}(\eta) \\ 0 \end{pmatrix} =  \dfrac{1}{\sqrt{2}} \begin{pmatrix} a v(\eta) \\ 0 \end{pmatrix}
\end{equation}
where we have defined $\tilde{v} = a v$.  For completeness, we discuss the equation of motion for $\tilde{v}$ in the Standard Model, including loop and finite temperature corrections, in Appendix \ref{ap:vhat_eqn_motion}.

\subsection{The Neutrino Sector}
\label{subsec:Neutrino_Sector}

Next, we consider the effect of the evolving Higgs VEV on the quadratic terms in $\tilde{\mathcal{L}}_\ell$, given by the second line of equations \eqref{eq:L_hats}.  Including multiple generations, we write this as
\begin{align}
\tilde{\mathcal{L}}_\ell &= i \sum_\alpha \overline{\tilde{L}_{\alpha}}\slashed{\partial} \tilde{L}_{\alpha} + i \sum_i \overline{\tilde{N}_{Ri}}\slashed{\partial} \tilde{N}_{Ri} -\sum_{\alpha \beta}  y_{\ell \alpha \beta} \bar{\tilde{L}}_{\alpha a} \tilde{\Phi}_{a} \tilde{\ell}_{\beta R} \nonumber \\
&- \sum_{\alpha j} y_{\nu \alpha j} \epsilon_{ab} \bar{\tilde{L}}_{\alpha a} \tilde{\Phi}_{b} \tilde{N}_{Rj}  - \sum_{ij} \dfrac{\tilde{M}_{Nij}}{2} \overline{(\tilde{N}_{Ri})^c} \tilde{N}_{Rj} + h.c.,
\label{eq:Lagrangian_1}
\end{align}
where $\tilde{\Phi}$ is the comoving Higgs doublet, $\tilde{L}$ is the comoving left-handed $(\nu_L, \ell_L)$ lepton $\mathrm{SU}_\mathrm{L}$ doublet of species $L$, and $\tilde{N}$ are right-handed Majorana neutrino states.  Greek indices label flavors ($e$, $\mu$, $\tau$), while the Latin indices $i$ and $j$ label right-handed neutrinos.  The indices $a$ and $b$ are $\mathrm{SU}_\mathrm{L}$ labels.  These are the only renormalizable terms which describe the interactions between the Higgs and lepton doublets, given the gauge symmetries of the Standard Model.

When the comoving Higgs field acquires a vacuum expectation value, this becomes:
\begin{align}
\tilde{\mathcal{L}}_\ell &= i \sum_\alpha \overline{\tilde{L}_{\alpha}}\slashed{\partial} \tilde{L}_{\alpha} + i \sum_i \overline{\tilde{N}_{Ri}}\slashed{\partial} \tilde{N}_{Ri}
-\sum_{\alpha \beta} \dfrac{y_{\ell \alpha \beta} \tilde{v}}{\sqrt{2}}  \overline{\tilde{\ell}_{L \alpha}} \tilde{\ell}_{R \beta}  \nonumber \\
&- \sum_{\alpha j}  \dfrac{y_{\nu \alpha j}\tilde{v}}{\sqrt{2}} \overline{\tilde{\nu}_{L \alpha}} \tilde{N}_{Ri} - \sum_{ij} \dfrac{\tilde{M}_{Nij}}{2} \overline{(\tilde{N}_{Ri})^c} \tilde{N}_{Rj} + h.c.,
\end{align}
where the comoving mass is $\tilde{M} = a M$.  The right-handed Majorana neutrinos induce lepton-number violation in interaction involving neutrinos; however, there is no corresponding effect for the charged leptons.  Therefore, these terms will not affect our analysis, and so we will absorb them into $\tilde{\mathcal{L}}_\mathrm{SM}$.  We define the neutrino sector Lagrangian
\begin{align}
\tilde{\mathcal{L}}_\nu &= i \sum_\alpha \overline{\tilde{\nu}_{L\alpha}}\slashed{\partial} \tilde{\nu}_{L\alpha} + i \sum_i \overline{\tilde{N}_{Ri}}\slashed{\partial} \tilde{N}_{Ri} - \sum_{\alpha i} \tilde{M}_{i \alpha}^D \overline{\tilde{N}_{Ri}} \tilde{\nu}_{L\alpha} \nonumber \\
&\qquad - \dfrac{1}{2} \sum_{ij} \tilde{M}_{Nij} \overline{(\tilde{N}_{Ri})^c} \tilde{N}_{Rj} + h.c., 
\label{eq:Lagrangian_flavor}
\end{align}
where the comoving Dirac mass is
\begin{equation}
\tilde{M}_{i \alpha}^D(\eta) = \dfrac{y_{\nu \alpha i}^\dagger \tilde{v}(\eta) }{\sqrt{2}}.
\end{equation}
We note that since $\tilde{v} = a v$, this has the expected scaling of a comoving mass.  

It will be convenient to use two-component comoving Weyl spinors; we work in the chiral basis, with conventions outlined in Appendix \ref{ap:Conventions}.  

We denote the two component spinors with tildes, as in
\begin{align}
\tilde{\nu}_L &= \begin{pmatrix}
\hat{\nu}_L \\ 0
\end{pmatrix}, \qquad \tilde{N}_R = \begin{pmatrix}
0 \\
\hat{N}_R
\end{pmatrix}, \nonumber \\
\overline{\tilde{\nu}_L} &= \begin{pmatrix} \hat{\nu}_L^\dagger & 0 \end{pmatrix} \begin{pmatrix} 0 & 1 \\ 1 & 0 \end{pmatrix} = \begin{pmatrix} 0 & \hat{\nu}_L^\dagger \end{pmatrix}, \nonumber \\
\overline{\tilde{N}_R} &= \begin{pmatrix} 0 & \hat{N}_R^\dagger \end{pmatrix} \begin{pmatrix} 0 & 1 \\ 1 & 0 \end{pmatrix} = \begin{pmatrix} \hat{N}_R^\dagger & 0 \end{pmatrix}.
\end{align}

The conjugated comoving fields are
\begin{align}
\overline{\tilde{N}_R^c} &= \overline{\mathcal{C} \overline{\tilde{N}_R}^T } = \begin{pmatrix} 0 & -i \hat{N}_R^T \sigma_2 \end{pmatrix},
\end{align}
where the charge conjugation operator $\mathcal{C}$ is also given in Appendix \ref{ap:Conventions}.

In terms of these two-component comoving spinors, the neutrino sector Lagrangian can be written as
\begin{align}
\tilde{\mathcal{L}}_\nu &= i \sum_\alpha \hat{\nu}_{L\alpha}^\dagger \bar{\sigma}^\mu \partial_\mu \hat{\nu}_{L\alpha} + i \sum_i \hat{N}_{Ri}^\dagger \sigma^\mu \partial_\mu \hat{N}_{Ri} \nonumber \\
&- \sum_{\alpha i} \left( \tilde{M}_{i \alpha}^D \hat{N}_{Ri}^\dagger \hat{\nu}_{L\alpha} + \tilde{M}_{\alpha i}^{D\dagger} \hat{\nu}_{L\alpha}^\dagger \hat{N}_{Ri} \right)  \nonumber \\
&  - \dfrac{1}{2} \sum_{ij} \left( i \tilde{M}_{Nij} \hat{N}_{Ri}^T \sigma_2 \hat{N}_{Rj} - i \tilde{M}_{Nji}^{\dagger} \hat{N}_{Rj}^\dagger \sigma_2 \hat{N}_{Ri}^* \right), 
\end{align}
and we note that if we introduce the fields
\begin{equation}
\hat{N}_R^C = -i \sigma_2 \hat{N}_R^*
\end{equation}
the mass term has the expected Majorana form,
\begin{equation}
-(\hat{N}_R^{C \dagger} \hat{N}_R + \hat{N}_R^\dagger \hat{N}_R^C),
\end{equation}
where we remind our readers that these are comoving fields.

\subsection{The $\mathcal{O}_6$ Operator}
\label{subsec:O6_Operator}

Next, we turn our attention to the $\mathcal{O}_6$ operator introduced in equation \eqref{eq:intro_O6}; we use the lepton and baryonic currents
\begin{align}
j^\mu_B &= \sum_{q}\dfrac{1}{n_c} q^\dagger \gamma^\mu q, \qquad
j^\mu_L = \sum_{\ell} \ell^\dagger \gamma^\mu \ell,
\end{align}
where the sums are over all leptonic fields, including right-handed neutrinos, and baryonic fields respectively.  In a general curved spacetime, we assume this becomes
\begin{align}
\tilde{\mathcal{L}}_{\mathcal{O}_6} &= -a^4 g^{\mu \nu} \dfrac{\Phi^2}{M^2} \nabla_\mu j_{\nu L},
\end{align}
which generally holds if the gravitational anomaly is cancelled by having equal numbers of left- and right-handed neutrinos, as discussed in Appendix \ref{ap:Generating_O_6_operator}.  Next, we integrate by parts by moving the derivative onto the Higgs vacuum expectation value.  As we are in flat FLRW spacetime, we may replace $\nabla_\mu$ with $\partial_\mu$ to find
\begin{align}
\tilde{\mathcal{L}}_{\mathcal{O}_6} &= -a^4 \dfrac{\partial_\mu \Phi^2}{M^2}  j^\mu_{B+L}.
\end{align}
Finally, we want to express this in terms of the comoving fields.  We note that the current is
\begin{align}
j_{\mu B+L} &= \bar{\psi} \tilde{\gamma}_\mu \psi 
= a^{-2} \tilde{j}_{\mu B+L},
\end{align}
where we have defined $\tilde{j}_{\mu B+L} = \bar{\tilde{\psi}} \gamma_\mu \tilde{\psi}$.  We here introduce the notation of a prime to denote a derivative with respect to conformal time.  This allows us to write
\begin{align}
\tilde{\mathcal{L}}_{\mathcal{O}_6} &= -\dfrac{\partial_\mu \tilde{\Phi}^2 - 2 a^\prime \tilde{\Phi}^2 \delta_{0 \mu} \slash a}{\tilde{M}^2} \tilde{j}^\mu_{B+L}.
\end{align}
where
\begin{align}
\partial_\mu \tilde{\Phi}^2 &= 2 a a^\prime \Phi^2 + a^2 \partial_\mu \Phi^2 \delta_{0 \mu} \nonumber \\
&= 2 \dfrac{a^\prime}{a} \tilde{\Phi}^2 \delta_{0 \mu} + a^2 \partial_\mu \Phi^2.
\end{align}

When $\tilde{\Phi}$ acquires a time-dependent vacuum expectation value $\tilde{v}$, this is
\begin{align}
\tilde{\mathcal{L}}_{\mathcal{O}_6} &= -\dfrac{\partial_0 \tilde{v}^2 - 2 a^\prime \tilde{v}^2 \slash a }{2 \tilde{M}^2} \tilde{j}^0_{B+L}.
\end{align}
where we emphasize that $\partial_0 = \partial \slash \partial \eta$.  We define
\begin{equation}
\tilde{\mu} \equiv \dfrac{\partial_0 \tilde{v}^2 - 2 a^\prime \tilde{v}^2 \slash a}{2 \tilde{M}^2}
\end{equation}
since this term acts like a chemical potential for $B+L$ charge.  We note that in terms of $v$ and $M$, this has the expected form
\begin{align}
\tilde{\mu} &\equiv -\dfrac{1}{2 \tilde{M}^2} \left( \dfrac{\partial}{\partial \eta} \tilde{v}^2 - 2 \dfrac{da}{d\eta} \dfrac{1}{a} \tilde{v}^2 \right) \nonumber \\
&= \dfrac{a}{M^2} v \dfrac{dv}{dt}.
\end{align}
Therefore, we also define
\begin{align}
\mu = -\dfrac{1}{M^2} v \dfrac{dv}{dt}
\end{align}
such that $\tilde{\mu} = a \mu$.  We note that since the Higgs VEV is initially decreasing, $dv \slash dt$ is initially negative.  The negative sign in $\mathcal{O}_6$ was chosen in order to bias the creation of particles over antiparticles.

As mentioned previously, lepton number is violated only in the neutrino sector, and therefore we are interested only in contribution to the current from the neutrinos.  Thus the relevant part of the $\mathcal{O}_6$ operator is
\begin{align}
\tilde{\mathcal{L}}_{\mathcal{O}_6} &= \tilde{\mu} \sum_\alpha \left( \begin{pmatrix}
0 & \hat{\nu}_{L \alpha}^\dagger \end{pmatrix} \begin{pmatrix} 0 & 1 \\ 1 & 0 \end{pmatrix} \begin{pmatrix} \hat{\nu}_{L \alpha} \\ 0 \end{pmatrix} \right. \nonumber \\
& \qquad \left.
+ \begin{pmatrix}
 \hat{N}_{R i}^\dagger & 0 \end{pmatrix} \begin{pmatrix} 0 & 1 \\ 1 & 0 \end{pmatrix} \begin{pmatrix} 0 \\ \hat{N}_{R i} \end{pmatrix}
 \right) \nonumber \\
&= \tilde{\mu} \sum_\alpha \left( \hat{\nu}_{L \alpha}^\dagger \hat{\nu}_{L \alpha} + \hat{N}_{R i}^\dagger \hat{N}_{R i} \right).
\label{eq:O6_relevent_part}
\end{align}

\subsection{Complete Two-Component Neutrino Lagrangian}

Using the results of the previous subsections, the complete effective Lagrangian for the comoving two-component neutrino fields is
\begin{align}
\tilde{\mathcal{L}} &= i \sum_\alpha \hat{\nu}_{L\alpha}^\dagger \bar{\sigma}^\mu \partial_\mu \hat{\nu}_{L\alpha} + i \sum_i \hat{N}_{Ri}^\dagger \sigma^\mu \partial_\mu \hat{N}_{Ri} \nonumber \\
&- \sum_{\alpha i} \left( \tilde{M}_{i \alpha}^D \hat{N}_{Ri}^\dagger \hat{\nu}_{L\alpha} + \tilde{M}_{\alpha i}^{D\dagger} \hat{\nu}_{L\alpha}^\dagger \hat{N}_{Ri} \right)  \nonumber \\
& - \dfrac{1}{2} \left( i \tilde{M}_{Nij} \hat{N}_{Ri}^T \sigma_2 \hat{N}_{Rj} - i \tilde{M}_{Nji}^\dagger \hat{N}_{Rj}^\dagger \sigma_2 \hat{N}_{Ri}^* \right) \nonumber \\
&+ \tilde{\mu} \sum_\alpha \left( \hat{\nu}_{L \alpha}^\dagger \hat{\nu}_{L \alpha} + \hat{N}_{R i}^\dagger \hat{N}_{R i} \right),
\label{eq:Full_Lagrangian}
\end{align}
which will be the basis for our subsequent analysis.  This describes a set of left-handed and right-handed neutrinos, with a Dirac mass and a right-handed Majorana mass, and a chemical potential for neutrino number, obtained from \eqref{eq:O6_relevent_part}.

\section{The Effective Lagrangian For One Generation of Left-Handed Neutrinos}
\label{sec:Effective_Lagrangian}

The Lagrangian in Eq.~\eqref{eq:Full_Lagrangian} includes several generations of both left and right-handed neutrinos.  It will be convenient to integrate out the heavy right-handed neutrinos\footnote{We do emphasize that we integrate out $\tilde{N}_R$, which are not, strictly speaking, identical with the heavy mass eigenstate.  This is a good approximation below the scale of the right-handed Majorana mass eigenvalues.} and specialize to a single generation, which will be a sufficiently rich model to capture the asymmetry production of interest here.

The comoving right-handed neutrinos obey the equations of motion
\begin{align}
0 &= i \sigma^\mu \partial_\mu \tilde{N}_{Ri}  - \sum_\alpha \hat{M}_{Di\alpha} \tilde{\nu}_{L\alpha} + i \sum_j (\hat{M}_{N}^\dagger)_{ij} \sigma_2 \tilde{N}_{Rj}^* \nonumber \\
& + \tilde{\mu} \tilde{N}_{Ri}.
\end{align}
In the limit of small $\tilde{\mu}$ and when the kinetic term is negligible, namely at scales below that of the right-handed Majorana mass eigenvalues, this equation is solved by
\begin{align}
\hat{N}_{Rk} &= - i \sum_{\alpha i} (\hat{M}_N^{T -1 })_{ki} (\hat{M}_{D}^*)_{i\alpha} \sigma_2 \tilde{\nu}_{L\alpha}^*,
\label{eq:eq_motion_N}
\end{align}
which when substituted into the Lagrangian gives 
\begin{widetext}
\begin{align}
\tilde{\mathcal{L}}_\mathrm{eff} 
&= i \sum_\alpha \hat{\nu}_{L\alpha}^\dagger \bar{\sigma}^\mu \partial_\mu \hat{\nu}_{L\alpha} - \dfrac{i}{2} \sum_{\alpha, \beta} \left[ (\tilde{M}_D^T \tilde{M}_N^{*-1} \tilde{M}_D)_{\alpha \beta} \hat{\nu}_{L\alpha}^T \sigma_2 \hat{\nu}_{L\beta} - (\tilde{M}_D^{\dagger} \tilde{M}_N^{T-1} \tilde{M}_D^*)_{\beta \alpha } \hat{\nu}_{L \beta}^\dagger \sigma_2 \hat{\nu}_{L \alpha }^*  \right]  + \tilde{\mu}  \sum_\alpha \hat{\nu}_{L \alpha}^\dagger \hat{\nu}_{L \alpha} \nonumber \\
& \qquad + \tilde{\mu} \sum_{\alpha, \beta} (\tilde{M}_D^\dagger \tilde{M}_N^{\dagger-1} \tilde{M}_N^{-1} \tilde{M}_D)_{\alpha \beta}  \hat{\nu}_{L \alpha}^\dagger \hat{\nu}_{L\beta} .
\end{align}

Doing so induces a Majorana mass for the left-handed neutrinos, of magnitude,
\begin{equation}
\tilde{M}_L = \tilde{M}_D^T \tilde{M}_N^{*-1} \tilde{M}_D.
\end{equation}
Thus this Lagrangian may be written (using implicit notation for the sums):
\begin{align}
\tilde{\mathcal{L}}_\mathrm{eff} &= i  \hat{\nu}_{L\alpha}^\dagger \bar{\sigma}^\mu \partial_\mu \hat{\nu}_{L\alpha} - \dfrac{i}{2}  \left[ (\tilde{M}_L)_{\alpha \beta} \hat{\nu}_{L\alpha}^T \sigma_2 \hat{\nu}_{L\beta}  - (\tilde{M}_L^\dagger)_{\beta \alpha } \hat{\nu}_{L \beta}^\dagger \sigma_2 \hat{\nu}_{L \alpha }^*  \right] + \tilde{\mu}  \hat{\nu}_{L \alpha}^\dagger \hat{\nu}_{L \alpha} + \tilde{\mu} (\tilde{M}_D^\dagger \tilde{M}_N^{\dagger-1} \tilde{M}_N^{-1} \tilde{M}_D)_{\alpha \beta}  \hat{\nu}_{L \alpha}^\dagger \hat{\nu}_{L\beta},
\end{align}
which has the equations of motion,
\begin{align}
0 &= i \bar{\sigma}^\mu \partial_\mu \hat{\nu}_{L\alpha} + i \sum_\beta (\tilde{M}_{L}^\dagger)_{\alpha \beta} \sigma_2 \hat{\nu}_{L \beta}^* + \tilde{\mu} \hat{\nu}_{L \alpha} + \tilde{\mu} (\tilde{M}_D^\dagger \tilde{M}_N^{\dagger-1} \tilde{M}_N^{-1} \tilde{M}_D)_{\alpha \beta}  \hat{\nu}_{L\beta}.
\end{align}
\end{widetext}


It is beneficial at this point to specialize to the one-generation case, since, as we will show, one generation is enough to obtain a nonvanishing asymmetry in the presence of the $\mathcal{O}_6$ operator.  We see explicitly that the induced Majorana mass transforms with the form appropriate to a comoving mass,
\begin{align}
\tilde{M}_L &= \dfrac{\tilde{M}_D^2}{\tilde{M}_N} = a \dfrac{y^2 v^2}{2 M_N}.
\end{align}
Thus we use the effective Lagrangian,
\begin{align}
\mathcal{L}_\mathrm{eff} &= i  \hat{\nu}_{L}^\dagger \bar{\sigma}^\mu \partial_\mu \hat{\nu}_{L} - \dfrac{i \tilde{M}_L}{2}  \left[ \hat{\nu}_{L}^T \sigma_2 \hat{\nu}_{L} - \hat{\nu}_{L }^\dagger \sigma_2 \hat{\nu}_{L  }^*  \right] + \tilde{\mu}_\mathrm{eff} \hat{\nu}_{L }^\dagger \hat{\nu}_{L },
\label{eq:Lagrangian}
\end{align}
where
\begin{align}
\tilde{\mu}_\mathrm{eff} = \tilde{\mu} \left( 1 + \dfrac{\tilde{M}_D^2}{\tilde{M}_N^2} \right)\approx \tilde{\mu},
\end{align}
wheen $\tilde{M}_D^2 \slash \tilde{M}_N^2 = y_\nu^2 v^2 \slash 2 M_N^2 \ll 1$.  We have rotated the field $\hat{\nu}_L$ to eliminate the phase in $\hat{M}_L$ which arises from the phase in the Yukawa coupling $y_\nu$.  (Note that the Higgs VEV $v$ can be taken to be real at all times.)

\section{Quantization and Bogoliubov Transformations}
\label{sec:Quantization}

Let us now discuss quantization.  First, we consider the scenario in which the mass and chemical potential are time-independent; we solve the equations of motion and determine the creation and annihilation operators which diagonalize the Hamiltonian.

Then we include the time-dependence of the mass and chemical potential, which induces a mixing between the positive and negative energy solutions of the field equation.  Consequently, even if the Hamiltonian is diagonal at time $t=0$, at a later time it will be non-diagonal.  It may be diagonalized with a time-dependent redefinition of the creation and annihilation operators; the coefficients of this diagonalization are known as the Bogoliubov coefficients, and in the subsequent section, we will relate these coefficients to the occupation number of physical eigenstates and to the lepton number.  This follows the procedure of e.g.~\cite{Mamaev:1976tq,Zeldovich:1971mw,Nilles:2001fg}.

We do note that in the multi-generation case, the time-dependent rotation that diagonalizes the mass matrix can introduce novel effects into particle production, as discussed in~\cite{Nilles:2001fg}; however, such features will not be necessary to generate a nonzero lepton number.  Therefore we work in the one generation limit, using Eq.~\eqref{eq:Lagrangian}, where these terms are absent.

\subsection{Constant Mass and Chemical Potential}
\label{subsec:Constant_M_Mu}

The equation of motion for Lagrangian with a single comoving Weyl field, Eq.~\eqref{eq:Lagrangian}, with constant comoving mass $\tilde{M}_L$ and comoving chemical potential $\tilde{\mu}_\mathrm{eff}$, is
\begin{equation}
(i \partial_0 - i \boldsymbol \sigma \cdot \boldsymbol \partial) \hat{\nu}_L = - \tilde{M}_L (i \sigma_2) \hat{\nu}_L^* - \tilde{\mu}_\mathrm{eff} \hat{\nu}_L.
\label{eq:eq_motion}
\end{equation}
or in momentum space,
\begin{equation}
(i \partial_0 + h |\tilde{\boldsymbol k}|) \hat{\nu}_L = - \tilde{M}_L (i \sigma_2) \hat{\nu}_L^* - \tilde{\mu}_\mathrm{eff} \hat{\nu}_L.
\end{equation} 
For consistency, we will use $\tilde{k}$ for the comoving momentum, and $p$ for the physical momentum.  We consider a solution of the form
\begin{align}
\hat{\nu}_L &= \int \dfrac{d^3\tilde{k}}{(2\pi)^{3}}  \sum_{h = \pm 1} \left[u(h,\tilde{k}) a^{(h)}_{\tilde{k}} \chi^{(h)}(\tilde{\boldsymbol k}) e^{i \tilde{\boldsymbol k} \cdot \boldsymbol x} \right. \nonumber \\
& \left. - v(h,\tilde{k})^* a^{(h)\dagger}_{\tilde{k}} \chi^{(-h)}(\tilde{\boldsymbol k}) e^{-i \tilde{\boldsymbol k} \cdot \boldsymbol x} \right],
\label{eq:nu_L_ansatz}
\end{align}
where $\chi^{(h)}(\tilde{\boldsymbol k})$ is the two-spinor which is an eigenstate of the helicity operator (appropriate to $\tilde{\boldsymbol k}$) with eigenvalue $h = \pm 1$.  This ansatz, when substituted into the equation of motion, requires
\begin{align}
(i \partial_0 + h |\tilde{\boldsymbol k}|) u(h,\tilde{k}) &= h \tilde{M}_L v(h,\tilde{k}) - \tilde{\mu}_\mathrm{eff} u(h,\tilde{k}) \nonumber \\
(i \partial_0 + h |\tilde{\boldsymbol k}|) v(h,\tilde{k})^* &= - h \tilde{M}_L u(h,\tilde{k})^* - \tilde{\mu}_\mathrm{eff} v(h,\tilde{k})^*.
\label{eq:us_and_vs_eqn_motion}
\end{align}
These equations can be decoupled,
\begin{equation}
(i \partial_0 + h |\tilde{\boldsymbol k}| + \tilde{\mu}_\mathrm{eff})(-i \partial_0 + h |\tilde{\boldsymbol k}| + \tilde{\mu}_\mathrm{eff}) F = - \tilde{M}_L^2 F
\end{equation}
with $F = u,v$, satisfies
\begin{equation}
\partial_0^2 F = \left[ (h|\tilde{\boldsymbol k}| + \tilde{\mu}_\mathrm{eff})^2 + \tilde{M}_L^2 \right] F.
\end{equation}
This has solutions of the form $F = e^{\pm i \tilde{\omega} \eta}$, where
\begin{align}
\tilde{\omega} \equiv \sqrt{(h|\tilde{\boldsymbol k}| + \tilde{\mu}_\mathrm{eff})^2 + \tilde{M}_L^2}.
\end{align}

Therefore, we take 
\begin{align}
u(h,\tilde{k}) &= \dfrac{\alpha}{\sqrt{2}} \sqrt{ 1 - f } e^{-i \tilde{\omega} \eta} + \dfrac{ \beta}{\sqrt{2}}     \sqrt{ 1 + f} e^{i \tilde{\omega} \eta}, \nonumber \\
v(h,\tilde{k}) &= \dfrac{h \alpha}{\sqrt{2}} \sqrt{ 1 + f } e^{-i \tilde{\omega} \eta}- \dfrac{h \beta}{\sqrt{2}} \sqrt{ 1 -f } e^{i \tilde{\omega} \eta},
\label{eq:new_soln}
\end{align}
where
\begin{equation}
f = \dfrac{h|\tilde{\boldsymbol k}| + \tilde{\mu}_\mathrm{eff}}{\tilde{\omega}} ,
\end{equation}
and $\alpha$ and $\beta$ are constant coefficients.  (In the time-dependent case, these will be the Bogoliubov coefficients.)  One can verify that these satisfy the first order equations of motion.

The state $\hat{\nu}_L$ obeys the anticommutation relations
\begin{align}
\left\lbrace \hat{\nu}_L(\boldsymbol x), \hat{\nu}_L(\boldsymbol y)^\dagger \right\rbrace &= \delta^{(3)}(\boldsymbol x - \boldsymbol y), \nonumber \\
\left\lbrace \hat{\nu}_L(\boldsymbol x), \hat{\nu}_L(\boldsymbol y) \right\rbrace &= 0, \qquad 
\left\lbrace \hat{\nu}_L^{\dagger}(\boldsymbol x), \hat{\nu}_L^{\dagger}(\boldsymbol y) \right\rbrace = 0.
\end{align}
These follow from the ansatz
\begin{align}
\left\lbrace a_{\tilde{k}}^{(h)}, a_{\tilde{q}}^{(\bar{h})\dagger} \right\rbrace &= (2\pi)^3 \delta^{(3)}(\tilde{\boldsymbol k} - \tilde{\boldsymbol q}) \delta_{h,\bar{h}}, \nonumber \\
\left\lbrace a_{\tilde{k}}^{(h)}, a_{\tilde{q}}^{(\bar{h})} \right\rbrace &= 0, \qquad
\left\lbrace a_{\tilde{k}}^{(h)\dagger}, a_{\tilde{q}}^{(\bar{h})\dagger} \right\rbrace = 0.
\end{align}
along with the normalization condition $|\alpha|^2 + |\beta|^2 = 1$.

Next we proceed to diagonalize the Hamiltonian; the appropriate creation and annihilation operators will not be the $a^{(h)}_{\tilde{\boldsymbol k}}$ and $a^{(h)\dagger}_{\tilde{\boldsymbol k}}$ operators themselves, but linear combinations of these operators.  Note that even with the $\mathcal{O}_6$ operator, the equation of motion ensures $\mathcal{L} = 0$; this follows from the fact that the equation of motion is first order.  Therefore the Hamiltonian is
\begin{align}
H = \dfrac{i}{2} \int d^3x  \left( \hat{\nu}_L^\dagger \partial_0 \hat{\nu}_L - (\partial_0 \hat{\nu}_L^\dagger) \hat{\nu}_L \right).
\label{eq:H_field}
\end{align}
In terms of the $a$ operators, this Hamiltonian is:
\begin{align}
H 
&= \dfrac{1}{2} \int \dfrac{d^3\tilde{k}}{(2\pi)^3} \sum_{h} \tilde{\omega} \left[ 2 \left[ |\alpha|^2 - |\beta|^2 \right]  a^{(h)\dagger}_{\tilde{k}} a^{(h)}_{\tilde{k}}   \right. \nonumber \\
& \left.+2  h \alpha^* \beta^* \zeta(\tilde{\boldsymbol k},h) a^{(h)\dagger}_{\tilde{k}}  a^{(h)\dagger}_{\tilde{k}_D} +2  h \alpha \beta \zeta(\tilde{\boldsymbol k},h)^* a^{(h)}_{\tilde{k}_D} a^{(h)}_{\tilde{k}}   \right],
\label{eq:H_a_operators}
\end{align}
where we have introduced the notation $p_D$ for the four-vector $(E,-\tilde{\boldsymbol k})$.  $\zeta(\tilde{\boldsymbol k},h)$ is a phase factor which arises from the product of the two spinors.  For the interested reader, the important steps in this derivation are discussed in Appendix \ref{ap:constant_mass_case}.  This can be written as a matrix equation:
\begin{align}
H &= \dfrac{1}{2} \int \dfrac{d^3\tilde{k}}{(2\pi)^3} \sum_{h} \tilde{\omega} 
\begin{pmatrix} a_{\tilde{k}}^{(h)\dagger} & a_{\tilde{k}_D}^{(h)} \end{pmatrix}
 \nonumber \\
& \qquad \cdot \begin{pmatrix} |\alpha|^2 - |\beta|^2 & 2  h \alpha^* \beta^* \zeta(\tilde{\boldsymbol k},h) \\
2  h \alpha \beta \zeta(\tilde{\boldsymbol k},h)^* & |\beta|^2 - |\alpha|^2 \end{pmatrix} \begin{pmatrix} a_{\tilde{k}}^{(h)} \\ a_{\tilde{k}_D}^{(h)\dagger} \end{pmatrix}
\end{align}
We introduce the rotated states:
\begin{align}
\begin{pmatrix}
A_{\tilde{k}}^{(h)\dagger} \\ A_{\tilde{k}_D}^{(h)}
\end{pmatrix} = \begin{pmatrix}
\alpha^* &  h \beta \zeta(\tilde{\boldsymbol k}, h)^* \\ -h \beta^* \zeta(\tilde{\boldsymbol k}, h) & \alpha
\end{pmatrix}  \begin{pmatrix}
a_{\tilde{k}}^{(h)\dagger}  \\ a_{\tilde{k}_D}^{(h)}
\end{pmatrix}
\label{eq:transformation_matrix}
\end{align}
%
which diagonalizes the Hamiltonian
\begin{align}
H &= \dfrac{1}{2} \int \dfrac{d^3\tilde{k}}{(2\pi)^3} \sum_{h} \tilde{\omega} 
(A_{\tilde{k}}^{(h)\dagger} A_{\tilde{k}}^{(h)} - A_{\tilde{k}_D}^{(h)} A_{\tilde{k}_D}^{(h)\dagger}) \nonumber \\
&=  \int \dfrac{d^3\tilde{k}}{(2\pi)^3} \sum_{h} \tilde{\omega} 
A_{\tilde{k}}^{(h)\dagger} A_{\tilde{k}}^{(h)},
\end{align}
where we have normal ordered and changed the integration variable to $-\boldsymbol p$ in the second term.  We note that since $h^2 = 1$, we can also write the eigenvalues as
\begin{align}
\tilde{\omega} \equiv \sqrt{(|\tilde{\boldsymbol k}| + h\tilde{\mu}_\mathrm{eff})^2 + \tilde{M}_L^2}.
\end{align}
Additionally, we note that as expected, $\tilde{\omega} = a \omega$, where 
\begin{align}
\omega = \sqrt{(|\boldsymbol p| + h\mu_\mathrm{eff})^2 + M_L^2},
\end{align}
and $\boldsymbol p$ is the physical momentum corresponding to the comoving momentum $\tilde{\boldsymbol k}$. 

\subsection{Time-Dependent Mass and Chemical Potential}
\label{subsec:Time_Dep_M_Mu}

Now we consider the case in which both the comoving mass and comoving chemical potential evolve in time.  We again use an expansion of the comoving Weyl spinor of the form of Eq.~\eqref{eq:nu_L_ansatz}, and the equation of motion again requires $u$ and $v$ to satisfy equations of the form of Eq.~\eqref{eq:us_and_vs_eqn_motion}, but with time dependent quantities $\tilde{M}_L$ and $\tilde{\mu}_\mathrm{eff}$.

We will consider solutions of the form
\begin{align}
u(h,\tilde{k}) &= \dfrac{\alpha}{\sqrt{2}} \sqrt{ 1 - f } e^{-i \int_0^\eta \tilde{\omega} d\bar{\eta}} + \dfrac{ \beta}{\sqrt{2}}     \sqrt{ 1 + f } e^{i \int_0^\eta \tilde{\omega} d\bar{\eta}} \nonumber \\
v(h,\tilde{k}) &= \dfrac{h \alpha}{\sqrt{2}} \sqrt{ 1 + f } e^{-i \int_0^\eta \tilde{\omega} d\bar{\eta}}- \dfrac{h \beta}{\sqrt{2}} \sqrt{ 1 - f } e^{i \int_0^\eta \tilde{\omega} d\bar{\eta}}.
\label{eq:new_soln_time_dep}
\end{align}
As shown in Appendix \ref{ap:alpha_beta_DEs}, this leads to the differential equations
\begin{align}
 \dfrac{d\alpha}{d\eta} &= - \dfrac{\beta}{2} \dfrac{ 1}{\tilde{\omega}^2} \left[ \tilde{M}_L \dfrac{d\tilde{\mu}_\mathrm{eff}}{d\eta} - (h |\tilde{\boldsymbol k}| + \tilde{\mu}_\mathrm{eff}) \dfrac{d\tilde{M}_L}{d\eta} \right] \nonumber \\
 & \qquad \cdot e^{2 i \int_0^\eta \tilde{\omega} d\bar{\eta}} \\
\dfrac{d\beta}{d\eta}&= \dfrac{\alpha}{2} \dfrac{ 1}{\tilde{\omega}^2} \left[ \tilde{M}_L \dfrac{d\tilde{\mu}_\mathrm{eff}}{d\eta} - (h |\tilde{\boldsymbol k}| + \tilde{\mu}_\mathrm{eff}) \dfrac{d\tilde{M}_L}{d\eta} \right] \nonumber \\
& \qquad \cdot e^{- 2 i \int_0^\eta \tilde{\omega} d\bar{\eta}}.
\label{eq:alpha_beta_DEs}
\end{align}
We take the initial conditions to be $\alpha(\eta=0) = 1$, $\beta(\eta=0)=0$.  This is consistent with the normalization condition $|\alpha|^2 + |\beta|^2 = 1$, and at $t=0$, the $A$ operators align with the $a$ operators, so the Hamiltonian (at this time) is diagonal when expressed in terms of either set.  The diagonalization of the Hamiltonian proceeds as in the time-independent case, as discussed in Appendix \ref{ap:alpha_beta_DEs}.

The effect of the time-dependent comoving mass $\tilde{M}_L$ and chemical potential $\tilde{\mu}_\mathrm{eff}$ is to mix positive and negative frequency modes, as is evident by the fact that $\beta$ will generally be nonzero at later times.  From the transformation matrix Eq.~\eqref{eq:transformation_matrix}, the operators that diagonalize the Hamiltonian at later times will generally be nontrivial linear combinations of $a^{(h)}_{\boldsymbol p}$ and $a^{(h)\dagger}_{\boldsymbol p}$.

\section{Particle Number and Lepton Number Operators}
\label{sec:Operators}

Next, we express the expectation values of the occupation number operator (for the physical eigenstates) and the lepton number operator in terms of the Bogoliubov coefficients $\alpha$ and $\beta$.  As the operators $A^{(h)}$ and $A^{(h)\dagger}$ diagonalize the Hamiltonian, these correspond to physical particles.    The procedure that we follow is this: we first express $N_h$ and $L_\mathrm{eff}$ in terms of these operators and normal order (for a discussion on normal ordering see~\cite{Barnaby:2012xt}).  We then express the operator in terms of the $a^{(h)}_{\boldsymbol p}$ and $a^{(h)\dagger}_{\boldsymbol p}$ operators using the transformation equations \eqref{eq:transformation_matrix}.  Then, we take the expectation value with the state $\left. | \mathrm{VAC},0 \right>$, the vacuum at time $t=0$.

The total number of physical particles of helicity $h$ is
\begin{align}
\tilde{N}_h &= \int \dfrac{d^3\tilde{k}}{(2\pi)^3} \left< \mathrm{VAC}; 0 | A_{\tilde{k}}^{(h)\dagger} A_{\tilde{k}}^{(h)} | \mathrm{VAC}; 0 \right>.
\end{align}
This operator is already normal-ordered, so we proceed to write this in terms of the time-independent $a_{\tilde{k}}^{(h)}$ operators,
\begin{align}
A_{\tilde{k}}^{(h)\dagger} A_{\tilde{k}}^{(h)} &= |\beta|^2 a_{\tilde{k}_D}^{(h)} a_{\tilde{k}_D}^{(h)\dagger} + h \alpha^*\beta^* \zeta(\tilde{\boldsymbol k},h) a_{\tilde{k}}^{(h)\dagger} a_{\tilde{k}_D}^{(h)\dagger} \nonumber \\
& \qquad + h \alpha \beta \zeta(\tilde{\boldsymbol k},h)^* a_{\tilde{k}_D}^{(h)} a_{\tilde{k}}^{(h)} + |\alpha|^2 a_{\tilde{k}}^{(h)\dagger} a_{\tilde{k}}^{(h)}.
\end{align}

We assume that we are in the state $\left. | \mathrm{VAC}, 0 \right>$, the vacuum state at time $t=0$.  Therefore, all operators of the form $a_{\tilde{k}}^{(h)}$ annihilate the vacuum.  Therefore,
\begin{align}
& \left< \mathrm{VAC};0 | A^{(h)\dagger}_{\tilde{k}} A^{(h)}_{\tilde{k}} | \mathrm{VAC};0 \right> \nonumber \\
&\qquad = |\beta|^2 \left< \mathrm{VAC}; 0 | a_{\tilde{k}_D}^{(h)} a_{\tilde{k}_D}^{(h)\dagger} | \mathrm{VAC}; 0 \right>  .
\end{align}
This matrix element is
\begin{small}
\begin{align}
&\left< \mathrm{VAC}; 0 | a_{\tilde{k}_D}^{(h)} a_{\tilde{k}_D}^{(h)\dagger} | \mathrm{VAC}; 0 \right> \nonumber \\
&= (2\pi)^3 \delta^{(3)}(0) \delta_{h,h} \left< \mathrm{VAC}; 0 | \mathrm{VAC};0 \right> 
 - \left< \mathrm{VAC};0| a_{\tilde{k}_D}^{(h)\dagger} a_{\tilde{k}_D}^{(h)} | \mathrm{VAC}; 0 \right> \nonumber \\
&=  V_{cm},
\end{align}
\end{small}
where $V_{cm}$ stands for the comoving volume (and we have used the usual formal manipulation $(2\pi)^3 \delta(0) = \int d^3 \tilde{x} e^{i \tilde{x} \cdot \boldsymbol 0} = V_{cm}$).  
This gives the expected result
\begin{align}
\tilde{N}_h &= V_{cm} \int \dfrac{d^3\tilde{k}}{(2\pi)^3} |\beta_{\tilde{\boldsymbol k},h}|^2,
\end{align}
where in general, $\beta$ may depend on the momentum and helicity, as we have noted.

Next we consider effective lepton number, which is carried by the neutrinos.  This charge is given by
\begin{align}
\tilde{L}_\mathrm{eff} &= \int d^3x \, \hat{\nu}_L^\dagger \hat{\nu}_L.
\label{eq:L_eff_def}
\end{align}
Following the procedure outline above gives us a normal ordered expression
\begin{widetext}
\begin{align}
\tilde{L}_\mathrm{eff} &= \int \dfrac{d^3\tilde{k}}{(2\pi)^{3 }} \sum_{h}  \left[ (-f) A_{\tilde{k}}^{(h)\dagger} A_{\tilde{k}}^{(h)} - \dfrac{\tilde{M}_L}{2 \tilde{\omega}} e^{2i \int_0^\eta \tilde{\omega} d\bar{\eta}}  \zeta(h,\tilde{\boldsymbol k}) A_{\tilde{k}}^{(h)\dagger} A_{\tilde{k}_D}^{(h)\dagger} - \dfrac{\tilde{M}_L}{2 \tilde{\omega}} e^{-2 i \int_0^\eta \tilde{\omega} d\bar{\eta}}  \zeta^*(h,\tilde{\boldsymbol k}) A_{\tilde{k}_D}^{(h)} A_{\tilde{k}}^{(h)} \right],
\label{eq:Leff_normal_ordered}
\end{align}
where the important steps are described in Appendix \ref{ap:L_eff}.  Taking the inner product with the $t=0$ vacuum gives
\begin{align}
&\left< \mathrm{VAC}; 0 | : \tilde{L}_\mathrm{eff}: | \mathrm{VAC}; 0 \right> = \int \dfrac{d^3\tilde{k}}{(2\pi)^{3 }} \sum_{h}  \left[ (-f) \left< \mathrm{VAC}; 0 | A_{\tilde{k}}^{(h)\dagger} A_{\tilde{k}}^{(h)} | \mathrm{VAC}; 0 \right>  \right. \nonumber \\
& \qquad \left. - \dfrac{\tilde{M}_L}{2 \tilde{\omega}} e^{2i \int_0^\eta \tilde{\omega} d\bar{\eta}}  \zeta \left< \mathrm{VAC}; 0 | A_{\tilde{k}}^{(h)\dagger} A_{\tilde{k}_D}^{(h)\dagger} | \mathrm{VAC}; 0 \right> - \dfrac{\tilde{M}_L}{2 \tilde{\omega}} e^{-2 i \int_0^\eta \tilde{\omega} d\bar{\eta}}  \zeta^* \left< \mathrm{VAC}; 0 | A_{\tilde{k}_D}^{(h)} A_{\tilde{k}}^{(h)} | \mathrm{VAC}; 0 \right> \right]
\end{align}
\end{widetext}
We express these in terms of the $a^{(h)}$ operators, which annihilate the state $\left. | \mathrm{VAC}; 0 \right>$; however, now that we have normal ordered we are careful to maintain any Dirac delta functions that arise from using the anticommutation relations.  The second and third matrix elements are
\begin{align}
\left< \mathrm{VAC}; 0 | A_{\tilde{k}}^{(h)\dagger} A_{\tilde{k}_D}^{(h)\dagger} | \mathrm{VAC}; 0 \right> 
&= h \alpha^* \beta \zeta^* V_{cm} \nonumber \\
\left< \mathrm{VAC}; 0 | A_{\tilde{k}_D}^{(h)} A_{\tilde{k}}^{(h)} | \mathrm{VAC}; 0 \right> 
&= h \alpha \beta^* \zeta V_{cm}
\end{align}
Therefore, the lepton number as a function of time is
\begin{align}
&\left< \mathrm{VAC}; 0 | : \tilde{L}_\mathrm{eff}: | \mathrm{VAC}; 0 \right> = V_{cm} \int \dfrac{d^3\tilde{k}}{(2\pi)^{3 }} \sum_{h}  \left[ (-f) |\beta^2| \right. \nonumber \\
& \left. - \dfrac{\tilde{M}_L}{2\tilde{\omega}} h \left(\alpha^* \beta e^{2i \int_0^\eta \tilde{\omega} d\bar{\eta}} + \alpha \beta^* e^{-2i \int_0^\eta \tilde{\omega} d\bar{\eta}} \right) \right]
\end{align}
where $f = (h |\tilde{\boldsymbol k}| + \tilde{\mu}_\mathrm{eff}) \slash \tilde{\omega}$.

\section{Rotated Operators}
\label{sec:Physical_Time}

At this point, it is convenient to define the rotated operators:
\begin{align}
\bar{\alpha} &= \alpha e^{-i \int_0^\eta \tilde{\omega} d\bar{\eta}} \nonumber \\
\bar{\beta} &= \beta e^{i \int_0^\eta \tilde{\omega} d\bar{\eta}}
\end{align}
which obey the differential equations
\begin{align}
\dfrac{d\bar{\alpha}}{d\eta} &=- c(\eta) \bar{\beta} - i \tilde{\omega} \bar{\alpha} \nonumber \\
\dfrac{d\bar{\beta}}{d\eta} &=  c(\eta) \bar{\alpha} + i \tilde{\omega} \bar{\beta},
\label{eq:rotated_DEs}
\end{align}
where 
\begin{equation}
c \equiv \dfrac{1}{2} \dfrac{ 1}{\tilde{\omega}^2} \left[ \tilde{M}_L \dfrac{d\tilde{\mu}_\mathrm{eff}}{d\eta} - (h |\tilde{\boldsymbol k}| + \tilde{\mu}_\mathrm{eff}) \dfrac{d\tilde{M}_L}{d\eta} \right],
\label{eq:c_def}
\end{equation}
and we also have the normalization condition $|\bar{\alpha}|^2 + |\bar{\beta}|^2 = 1$ along with the initial condition $\bar{\alpha}=1$ and $\bar{\beta}=0$.  

Note that we can rewrite this so that the helicity $h$ multiplies the chemical potential,
\begin{equation}
c = \dfrac{h}{2} \dfrac{ 1}{\tilde{\omega}^2} \left[ \tilde{M}_L h\dfrac{d\tilde{\mu}_\mathrm{eff}}{d\eta} - ( |\tilde{\boldsymbol k}| + h \tilde{\mu}_\mathrm{eff}) \dfrac{d\tilde{M}_L}{d\eta} \right].
\end{equation}

In terms of these rotated coefficients, the number densities and lepton number are:
\begin{align}
&\left< \mathrm{VAC}; 0 | :\tilde{N}_h: | \mathrm{VAC}; 0 \right> = V_{cm} \int \dfrac{d^3\tilde{k}}{(2\pi)^3} |\bar{\beta}_{\tilde{\boldsymbol k},h}|^2, \nonumber \\
&\left< \mathrm{VAC}; 0 | : \tilde{L}_\mathrm{eff}: | \mathrm{VAC}; 0 \right> = V_{cm} \int \dfrac{d^3\tilde{k}}{(2\pi)^{3 }} \sum_{h} \nonumber \\
&  \left[ - \dfrac{h |\tilde{\boldsymbol k}| + \tilde{\mu}_\mathrm{eff}}{\tilde{\omega}} |\bar{\beta}^2_{\tilde{\boldsymbol k},h}| - \dfrac{\tilde{M}_L}{2\tilde{\omega}} h \left(\bar{\alpha}^*_{\tilde{\boldsymbol k},h} \bar{\beta}_{\tilde{\boldsymbol k},h}  + \bar{\alpha}_{\tilde{\boldsymbol k},h} \bar{\beta}_{\tilde{\boldsymbol k},h}^* \right) \right]
\label{eq:operators_rotated_coeffs}
\end{align}
which has eliminated the fast oscillatory time dependence.  
The comoving number densities are therefore
\begin{align}
&\tilde{n}_h  =  \int \dfrac{d^3\tilde{k}}{(2\pi)^3} |\bar{\beta}_{\tilde{\boldsymbol k},h}|^2, \nonumber \\
&\tilde{n}_L = \int \dfrac{d^3\tilde{k}}{(2\pi)^{3 }} \sum_{h} \nonumber \\
&  \left[ - \dfrac{h |\tilde{\boldsymbol k}| + \tilde{\mu}_\mathrm{eff}}{\tilde{\omega}} |\bar{\beta}_{\tilde{\boldsymbol k},h}|^2 - \dfrac{\tilde{M}_L}{2\tilde{\omega}} h \left(\bar{\alpha}^*_{\tilde{\boldsymbol k},h} \bar{\beta}_{\tilde{\boldsymbol k},h}  + \bar{\alpha}_{\tilde{\boldsymbol k},h} \bar{\beta}_{\tilde{\boldsymbol k},h}^* \right) \right]
\end{align}
At late times, the Higgs VEV $v$ approaches zero, and therefore the comoving VEV $\tilde{v} = a v$ also does.  Consequently, $\tilde{M}_L \rightarrow 0$, $\tilde{\mu}_\mathrm{eff} \rightarrow 0$, and $\tilde{\omega} \rightarrow |\tilde{\boldsymbol k}|$.  Therefore, the limit of the comoving lepton asymmetry is
\begin{align}
\lim_{t \rightarrow \infty} \tilde{n}_L = \int \dfrac{d^3\tilde{k}}{(2\pi)^{3 }} \sum_{h} (-h)  |\bar{\beta}_{\tilde{\boldsymbol k},h}|^2 .
\end{align}
As expected, this is the difference in the number of helicity states.

The physical number density and lepton density are
\begin{align}
n_h &= \dfrac{1}{a(t)^3} \int \dfrac{d^3\tilde{k}}{(2\pi)^3} |\bar{\beta}_{\tilde{\boldsymbol k},h}|^2, \nonumber \\
\lim_{t \rightarrow \infty} n_L &= \dfrac{1}{a(t)^3} \int \dfrac{d^3\tilde{k}}{(2\pi)^{3 }} \sum_{h} (-h)  |\bar{\beta}_{\tilde{\boldsymbol k},h}|^2.
\end{align}
The final lepton asymmetry is given by
\begin{align}
\eta_L &\equiv \lim_{t \; \mathrm{large}} \dfrac{n_L}{n_\gamma} \nonumber \\
&= -\dfrac{\pi}{2 \zeta(3) T(t)^3} \dfrac{1}{a(t)^3} \int \dfrac{d^3\tilde{k}}{(2\pi)^{3 }} \sum_{h} (-h)  |\bar{\beta}_{\tilde{\boldsymbol k},h}|^2,
\label{eq:eta_final}
\end{align}
which should be evaluated at a time after the completion of reheating, so that $a(t) T(t)$ approaches an asymptotic constant value, but before electroweak sphalerons redistribute the charge between lepton and baryons.  Following this, further entropy production results in a final baryonic asymmetry about an order of magnitude smaller than $\eta_L$.

\section{Approximations and Numerical Analysis}
\label{sec:Numerics}

During the evolution of the Higgs VEV, $c(\eta)$ (defined in \eqref{eq:c_def}) is nonzero, which results in $\bar{\beta}(\eta) \neq 0$, signaling particle production.  Additionally, since $c(\eta,h=+) \neq c(\eta,h=-)$ generically, we expect a nonzero lepton asymmetry.  At late times, $c(\eta) \rightarrow 0$ for both helicity values, resulting in $\bar{\beta} \sim \exp( i |\tilde{\boldsymbol k}| \eta)$, which gives a nonzero asymptotic value for $|\tilde{\beta}|^2$, which is not generically identical for the two helicity values.   Therefore, we expect a nonzero asymmetry to survive at late times, after the Higgs VEV (and hence $M_L$ and $\mu$) approaches zero.  

Calculating this asymmetry is complicated by the lack of analytic closed form solutions  to the differential equations \eqref{eq:rotated_DEs}, which must be solved numerically.  In this section, we introduce a sequence of useful approximations which simplify this problem significantly; we then present a numerical analysis of the resulting asymmetry.  We focus particularly on the range of parameter space in which an asymmetry matching the observed cosmological baryonic abundance is generated.

\subsection{Higgs Oscillations}
\label{subsec:Higgs_VEV}

We first note that it is desirable to have significant damping in the oscillations of the Higgs VEV.  This is because the chemical potential $\tilde{\mu}_\mathrm{eff} \sim v v^\prime$ changes sign frequently during the oscillations, and so whether particle or antiparticle production is favored also oscillates.  Therefore, a significant damping in the amplitude of the oscillation avoids washout from this alternation.

As we explain below, this allows us to make two simplifications: First, that the asymmetry production occurs on a time scale during which $a(t)$ is approximately constant, and second, particle production occurs primarily in those comoving momenta least affected by washout.

We noted above that washout is significant unless the Higgs VEV is significantly damped.  Consequently, the asymmetry production is dominated by the particle production during the initial relaxation of the Higgs VEV, which may be a fast process, compared to the evolution of the universe.  We have mentioned in Sec.~\ref{subsec:Higgs_Sector} that there are several reasons why the Higgs field may have a large vacuum expectation value after inflation.  In one scenario, the Higgs VEV grows due to quantum fluctuations within the unmodified Standard Model, or alternatively, the Higgs field may be trapped in a false vacuum during inflation.  In the latter example, it is quite natural that the evolution of the Higgs VEV, once it is released from the false vacuum, would occur on time scales $\tau \ll 1 \slash H$.  This is more difficult to arrange in the former scenario, as the condition for the VEV to grow requires $m_\mathrm{eff} \lesssim H_I$, and the time scale of the Higgs VEV evolution is $\sim 1 \slash m_\mathrm{eff}$.     Rapid evolution of the Higgs VEV may still be arranged, as both $m_\mathrm{eff}$ and $H$ are functions of time, although this may be somewhat unnatural.

In the limit that the evolution of the Higgs VEV is rapid compared to the expansion of the universe, we may approximate $a(t)\sim a(t_S)$ constant, where we define $t=t_S$ to be the time at which the Higgs VEV begins rolling significantly.  The comoving momentum during the epoch of particle production is $\tilde{\boldsymbol k} = a(t_S) \boldsymbol p$, and it is convenient to express $\bar{\beta}$ and $\bar{\alpha}$ as functions of physical time $t$ instead of conformal time $\eta$.  They obey the differential equations
\begin{align}
\dfrac{d\bar{\alpha}}{dt} &=- C(t) \bar{\beta} - i \omega \bar{\alpha}, \nonumber \\
\dfrac{d\bar{\beta}}{dt} &=  C(t) \bar{\alpha} + i \omega \bar{\beta},
\label{eq:DEs_tilde}
\end{align}
with
\begin{align}
C(t) = \dfrac{c(t)}{a(t)} &= \dfrac{h}{2 \omega^2} \left[ M_L \left( H(t) \mu_\mathrm{eff} + \dot{\mu}_\mathrm{eff}\right) \right. \nonumber \\
& \left. - \left( |\boldsymbol p| + h \mu_\mathrm{eff} \right)^2 \left( H(t) M_L + \dot{M}_L \right) \right],
\end{align}
We remind our readers that the untilded $\mu_\mathrm{eff}$, $M_L$, and $\omega$ are the physical, and not comoving, quantities.  For self consistency, we drop the terms proportional to $H(t)$, giving
\begin{align}
C(t) &\approx \dfrac{h}{2 \omega^2} \left[  M_L   \dot{\mu}_\mathrm{eff} - \left( |\boldsymbol p| + h \mu_\mathrm{eff} \right)  \dot{M}_L  \right].
\label{eq:c_slash_a}
\end{align}
We emphasize that these expressions involve the physical, not comoving, momentum.  However, when the evolution of the Higgs VEV is fast, these are related by the constant factor $a(t_S)$; we use this assumption to write
\begin{align}
\eta_L 
&= -\dfrac{\pi a(t_S)^3}{2 \zeta(3) T(t)^3 a(t)^3} \int \dfrac{d^3p}{(2\pi)^{3 }} \sum_{h} (-h)  |\bar{\beta}_{\boldsymbol p,h}(t_E)|^2,
\end{align}
where $t_E$ is the effective end of particle production.  We emphasize that our assumption is that $a(t)$ is approximately constant while the neutrino asymmetry is produced (for $t_S \leq t \leq t_E $), which allows us to use $|\tilde{k}| \sim a(t_S) |\boldsymbol p|$ in the integral of Eq.~\eqref{eq:eta_final}.  Once the Higgs VEV relaxes to zero, no further asymmetry is produced; however, the physical volume continues expanding $\sim a(t)^3$.  This is responsible for the factor of $a(t)^3$ in the denominator, which may be large; that is, this equation continues to hold even when $a(t) \slash a(t_S) \gg 1$, provided that $a(t_E) \sim a(t_S)$.

\begin{figure}
\begin{center}
\includegraphics[scale=.4]{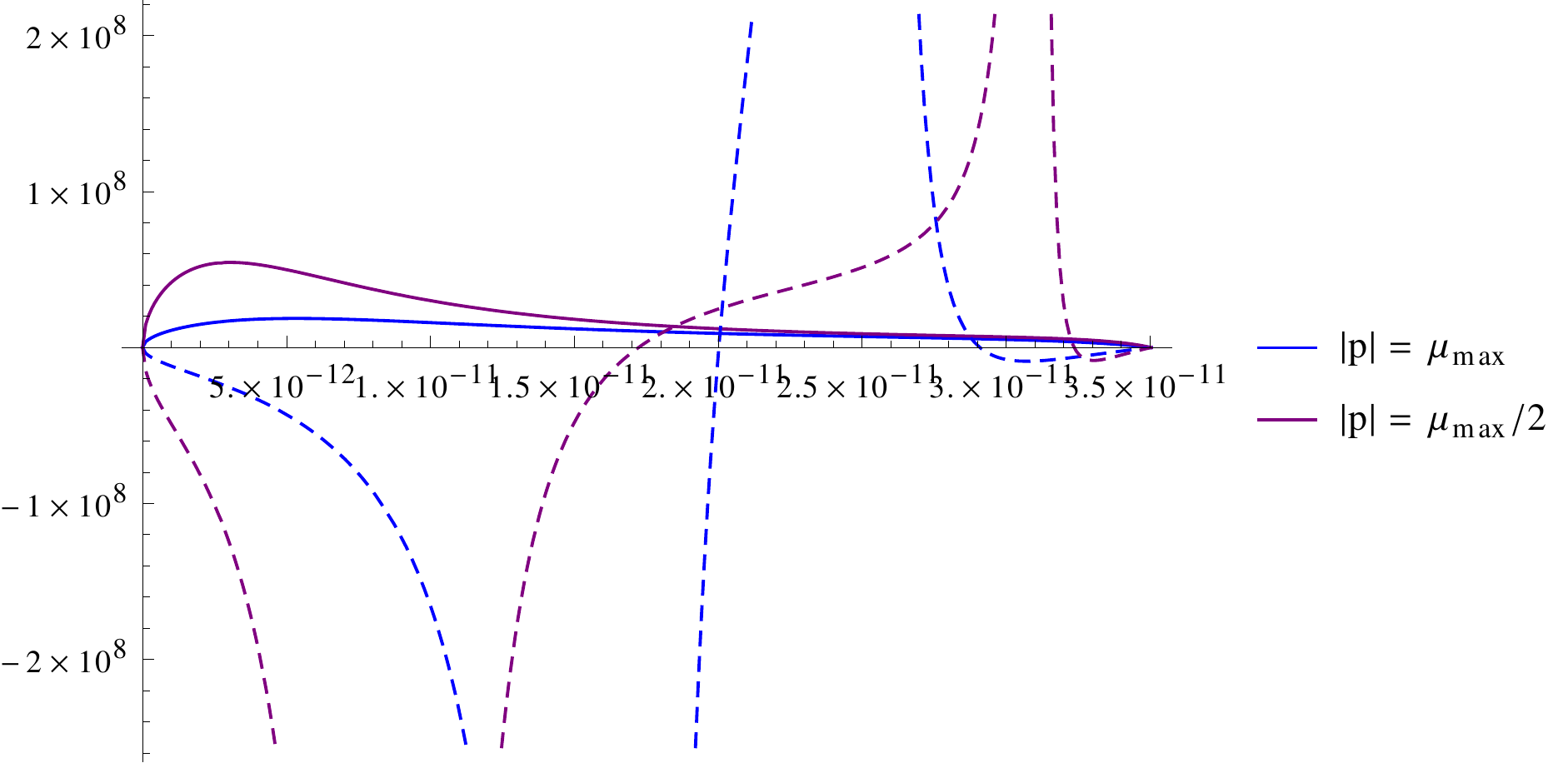}
\includegraphics[scale=.4]{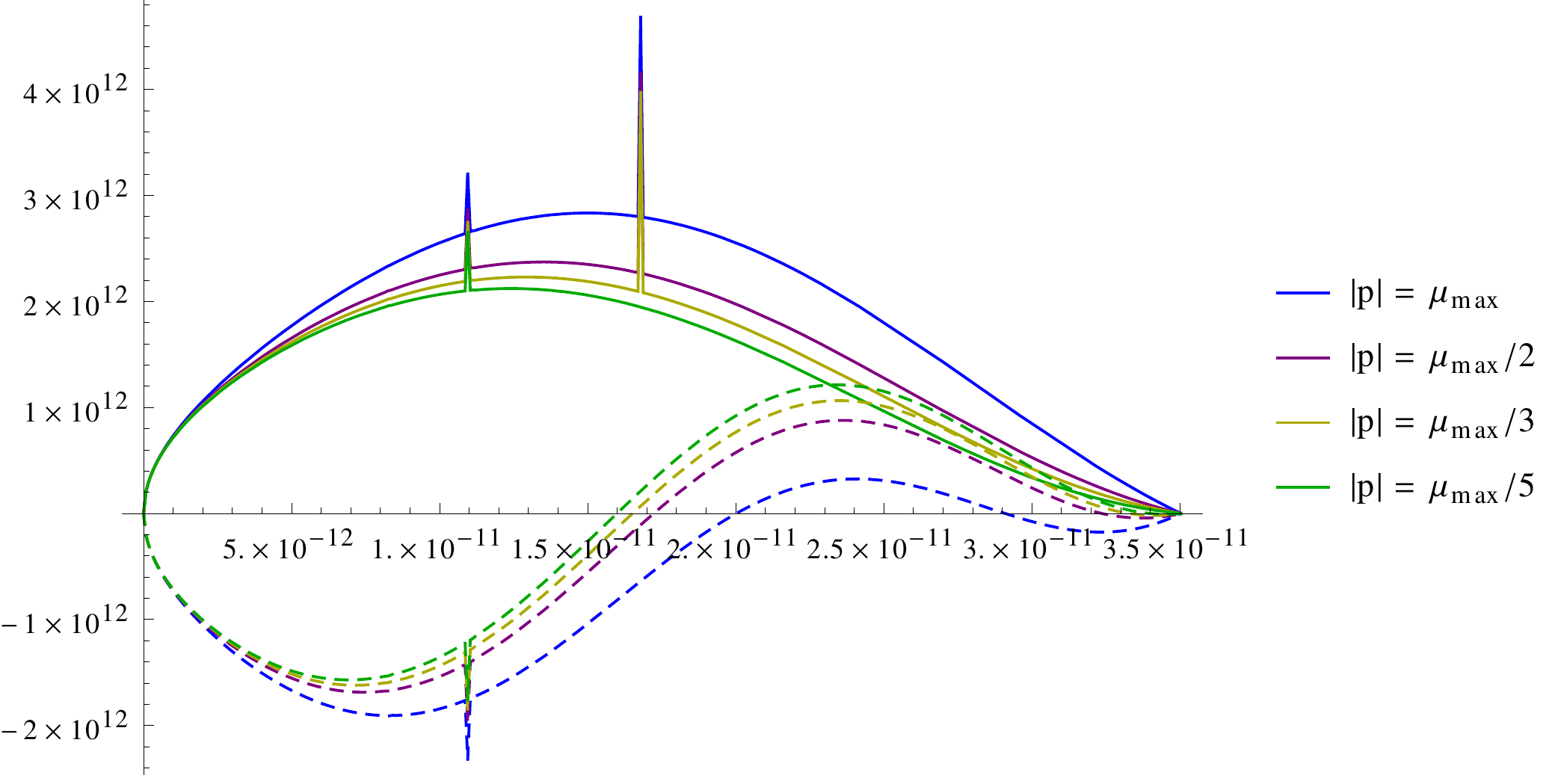}
\end{center}
\caption{$C(t)$ (top) and $\omega(t)^2 C(t)$ (bottom).  The solid lines are for $h = + 1$, while the dashed lines are for $h = -1$.  These plots cover the initial pass of the Higgs VEV to zero.  For concreteness, we have used the IC-2 scenario of Ref.~\cite{Kusenko:2014lra,Yang:2015ida} with the parameters $\Lambda_I = 10^{13}$~GeV and $\Gamma_I = 10^5$~GeV, along with $M$, the scale in $\mathcal{O}_6$, equal to $10^{11}$~GeV.  This is about four orders of magnitude larger than the initial Higgs VEV.  $\mu_\mathrm{max}$ is the maximum of $|\mu(t)|$.  Units are appropriate powers of GeV.} 
\label{fig:impossible}
\end{figure}

Next we observe that if $|\bar{\beta}|^2 \ll 1$ at all times, we can approximate $\bar{\alpha} \approx 1$ and the relevant differential equation is simply
\begin{align}
\bar{\beta} &= \int_0^t C(\bar{t}) d\bar{t}.
\label{eq:approx_beta_DE}
\end{align}
Sample plots of $\omega(t)^2 C(t)$ (the factor $\omega^2$ cancels the $1 \slash \omega^2$ dependence in Eq.~\eqref{eq:c_slash_a}) and $C(t)$ are shown in Fig.~\ref{fig:impossible}.  We note that $M_L \slash \mu_\mathrm{eff} \sim y_\nu^2 M^2 v \slash M_N \dot{v} \sim y_\nu^2 M^2 \slash H_I M_N$, and $y_\nu^2 \slash M_N \sim 10^{-20} \; \mathrm{GeV}^{-1}$ is fixed by the observed neutrino masses differences.  Therefore, it is not surprising that for these parameters the typical scale of $\mu_\mathrm{eff} \approx \mu$ is about 12 orders of magnitude larger than the typical scale of $M_L$.  We proceed to describe the qualitative behavior of these plots.  

The $C(t)$ panel (top) has a sequence of sharp peaks for $h = -1$, of alternating sign, while these peaks are absent for $h = +1$.  These are a consequence of the $1 \slash \omega^2$ factor in $C(t)$.  We note that during the initial pass of the Higgs field towards zero, $\mu(t)$ is positive, and therefore, $\omega \sim |\boldsymbol p| + h \mu$ has a significant cancellation when $|\boldsymbol p| \sim \mu$ for $h = -1$, while for $h=+1$ these factors always add.  Additionally, for $h=-1$, $|\boldsymbol p| + h \mu$ changes sign at this peak.  As the scale of $\dot{\mu}$ is larger than the scale of the left-handed neutrino mass $\dot{M}_{L}$, $C(t)$ changes sign ``within" this peak.  This is responsible for the very sharp positive and negative peaks.  When evaluating this integral, these peaks cancel to a significant precision.

To understand the behavior better, we consider $\omega^2 C(t)$, which eliminates the sharp peaks.  This plot is shown in the bottom panel of Fig.~\ref{fig:impossible}.   We observe that in contrast to the $C(t)$ plot, the $h=+1$ functions generally have smaller magnitudes than their $h=-1$ counterparts.  This is because $C(t) \propto |\boldsymbol p| + h \mu$, and so the is a partial cancellation that affects $\omega$ also affects the overall magnitude of $C(t)$ for $h = -1$.

Additionally, for $h = -1$, the factors of $(|\boldsymbol p| + h \mu)$ changes sign (since $\mu$ is negative), which affects the $ \dot{M}_L$.  This factor is responsible for the various sign changes in the bottom plot of Fig.~\ref{fig:impossible}, even though the Higgs VEV is decreasing during the entire time shown.  We also note that the sharp spikes occur when $\dot{\mu} = 0$, and so, momentarily, these plots are dominated by the $\dot{M}_{L}$ term, which (at this time) happens to be much larger in magnitude.

As a simpler toy model, we consider a Higgs field which obeys the equation of motion,
\begin{align}
\dfrac{d^2v}{dt^2} + 3 H \dfrac{dv}{dt} + m^2 v + \Gamma_H \dfrac{dv}{dt} = 0,
\end{align}
along with the boundary condition $v(t_0) = v_0$, $\dot{v}(t_0) = 0$.  This is easier to analyze numerically, as opposed to considering the Higgs potential with running, and temperature-dependent, coupling constants.  Furthermore, the Standard Model Higgs field decays primarily through non-perturbative effects~\cite{Enqvist:2013kaa,Figueroa:2015rqa}.  For self-consistency, we again assume the Hubble friction term is negligible.  This has the approximate solution
\begin{align}
v(t) = v_0 e^{- \Gamma_H (t-t_0) \slash 2} \cos(\Omega (t-t_0)),
\end{align}
where $\Omega = \sqrt{m^2 - \Gamma_H^2}$.  We have taken $t=0$ as the time at which the Higgs starts oscillating, and so $t_S = 0$.

We note that a particularly interesting scenario is the case in which the potential $V(v) = \lambda v^4 \slash 4$; this is well-motivated by the fact that the Standard Model potential at large VEVs is dominated by this term.  In this case, the term $m^2 v$ in the equation of motion would be replaced by $\lambda_\mathrm{eff} v^3$, and in the solution the cosine function would instead be $\mathrm{cn}(v_0 t \slash \lambda^{1 \slash 4})$, where $\mathrm{cn}$ is a Jacobi sinusoidal function~\cite{Frasca:2009yp}.  

The Higgs potential is not known at large VEVs; therefore, we will consider the parameters $v_0$, $m$, and $\Gamma_H$ to be independent, and furthermore, which may be chosen independently of any parameters describing inflation and reheating.  We do note that in the scenario in which the initially large Higgs field VEV is produced via quantum fluctuations, $v_0$ will be determined by the scale of inflation, although it will also be affected by any higher dimensional operators that influence the Higgs potential.

We also made these further approximations: First, we assumed $\Omega \gg \Gamma_H$, such that
\begin{align}
M_L &\approx \dfrac{y^2 v_0^2}{4 M_N} e^{-\Gamma_H t } \left( 1 - \cos(2 \Omega t) \right),  \nonumber \\
\dot{M}_{L} &\approx - \dfrac{y^2 v_0^2 \Omega}{2 M_N} e^{-\Gamma_H t } \sin(2 \Omega t), \nonumber \\
\mu &\approx \dfrac{v_0^2 \Omega}{2 M^2} e^{-\Gamma_H t } \sin(2 \Omega t), \nonumber \\
\dot{\mu} &\approx \dfrac{v_0^2 \Omega^2}{M^2} e^{-\Gamma_H t } \cos(2 \Omega t).
\end{align}
Although we want significant damping, this is self-consistent, as we must have $\Omega \gtrsim \Gamma_H$ in order for the Higgs VEV to undergo oscillatory motion.

Next, we observe that the $d^3p$ integral is dominated by momenta $|\boldsymbol p| \sim \mu_\mathrm{max}$, where $\mu_\mathrm{max}$ is the maximum of $|\mu(t)|$  This is because these momentum values suffer the least washout during the subsequent oscillations of the Higgs VEV.  Therefore, we approximate
\begin{align}
\eta &\approx -\dfrac{\pi}{2 \zeta(3) T(t)^3} \dfrac{a(t_S)^3}{a(t)^3} \dfrac{\mu_\mathrm{max}^3}{(2\pi)^{3 }} \sum_{h} h |\bar{\beta}_{\mu_\mathrm{max},h}(t_E)|^2.
\label{eq:eta_after_higgs_approx}
\end{align}

\subsection{Low Scale of Inflation}

For simplicity, we assume coherent oscillations begin instantly at the end of the inflationary epoch.  We normalize the scale factor to one at the end of inflation, when the coherent oscillations of the inflaton start.  In the following computations, we approximate that the higgs oscillations also start at this time.  This is a good approximation in the case when the Higgs VEV is prevented from rolling by Hubble friction; when the Higgs is instead trapped in a false vacuum, this approximation will only be valid for sufficicently small barriers.  

Next we consider the factor of $a(t)^3 T(t)^3$ in the denominator of $\eta$ given by Eq.~\eqref{eq:eta_after_higgs_approx}.  We emphasize that $a(t) T(t)$ is the value approached at relatively late times, well into the radiation dominated epoch after Higgs relaxation has ended, but before the Standard Model degrees of freedom have decoupled.  This is completely determined by the two inflationary parameters, the inflationary scale $\Lambda_I$ and the decay rate of the inflaton, $\Gamma_I$, where the Hubble parameter during inflation is
\begin{equation}
H_I = \sqrt{ \dfrac{8\pi}{3}} \dfrac{\Lambda_I^2}{M_\mathrm{Pl}}.
\end{equation}
Since we do not fix a specific model of inflation, we take these to be independent parameters.  We note that $a(t) T(t)$ becomes constant once reheating has completed, and the asymptotic value is shown in Figure \ref{fig:aT_plot}.  We can obtain the scaling of this factor as a function of the parameters of the inflationary sector, under the assumption of instantaneous inflaton decay at $H = \Gamma_I$ and of instantaneous thermalization. This gives:

\begin{figure}
\begin{center}
\includegraphics[scale=.6]{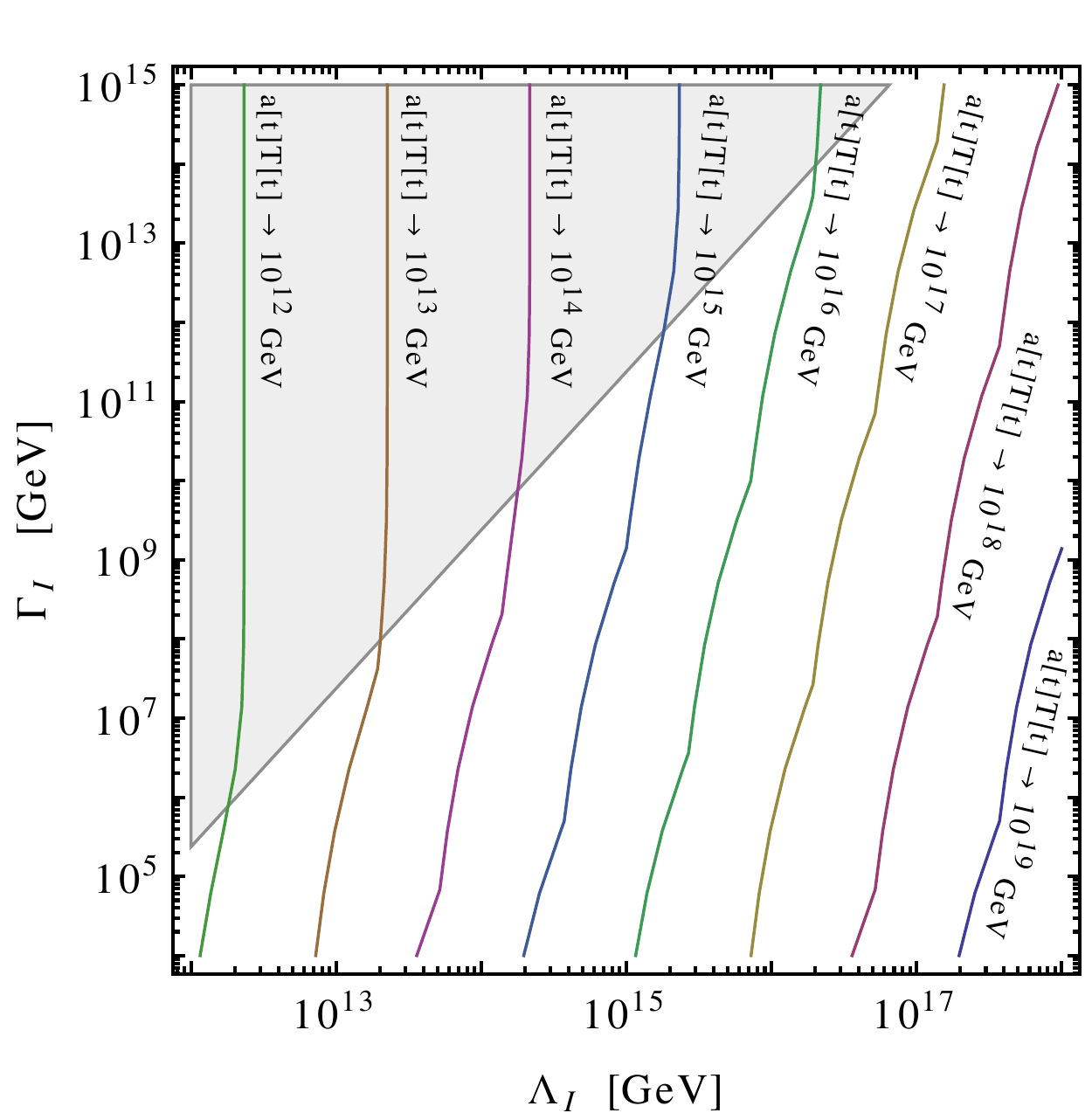}
\caption{Contours of asymptotic late time $aT$; the scale factor is normalized to 1 at the time the inflaton starts oscillating.  In the gray area, $\Gamma_I > H_I$, and there is no inflationary epoch.}
\label{fig:aT_plot}
\end{center}
\end{figure}

\begin{align}
a(t) T(t) \propto \dfrac{\Lambda_I^{4 \slash 3}}{M_\mathrm{Pl}^{1\slash 6} \Gamma_I^{1 \slash 6}},
\end{align}
or in terms of the reheat temperature,
\begin{align}
\dfrac{1}{(a(t)T(t))^3} \propto \dfrac{T_{RH}}{\Lambda_I^4}.
\end{align}
where we recall that the scale factor $a$ has been normalized to one at the end of inflation, which we take to be simultaneous with the beginning of Higgs relaxation.  We see that a large baryon asymmetry is obtained for a low inflationary scale. For this reason we consider values in the bottom left corner of Fig.~\ref{fig:aT_plot}, which are characterized by a relatively low $\Lambda_I$. By consistency, $T_{RH}$ then needs to be below $\Lambda_I$.

We note that altering $\Lambda_I$ and $\Gamma_I$ may modify the evolution of the Higgs VEV, particularly if finite temperature corrections to the Higgs potential are significant.  In particular, in scenarios in which the Higgs VEV begins in a false minimum which is destabilized by thermal fluctuations, there is a minimum reheat temperature which constrains $\Lambda_I$ and $\Gamma_I$.

\subsection{Numerical Example}

\begin{figure}
\includegraphics[scale=.4]{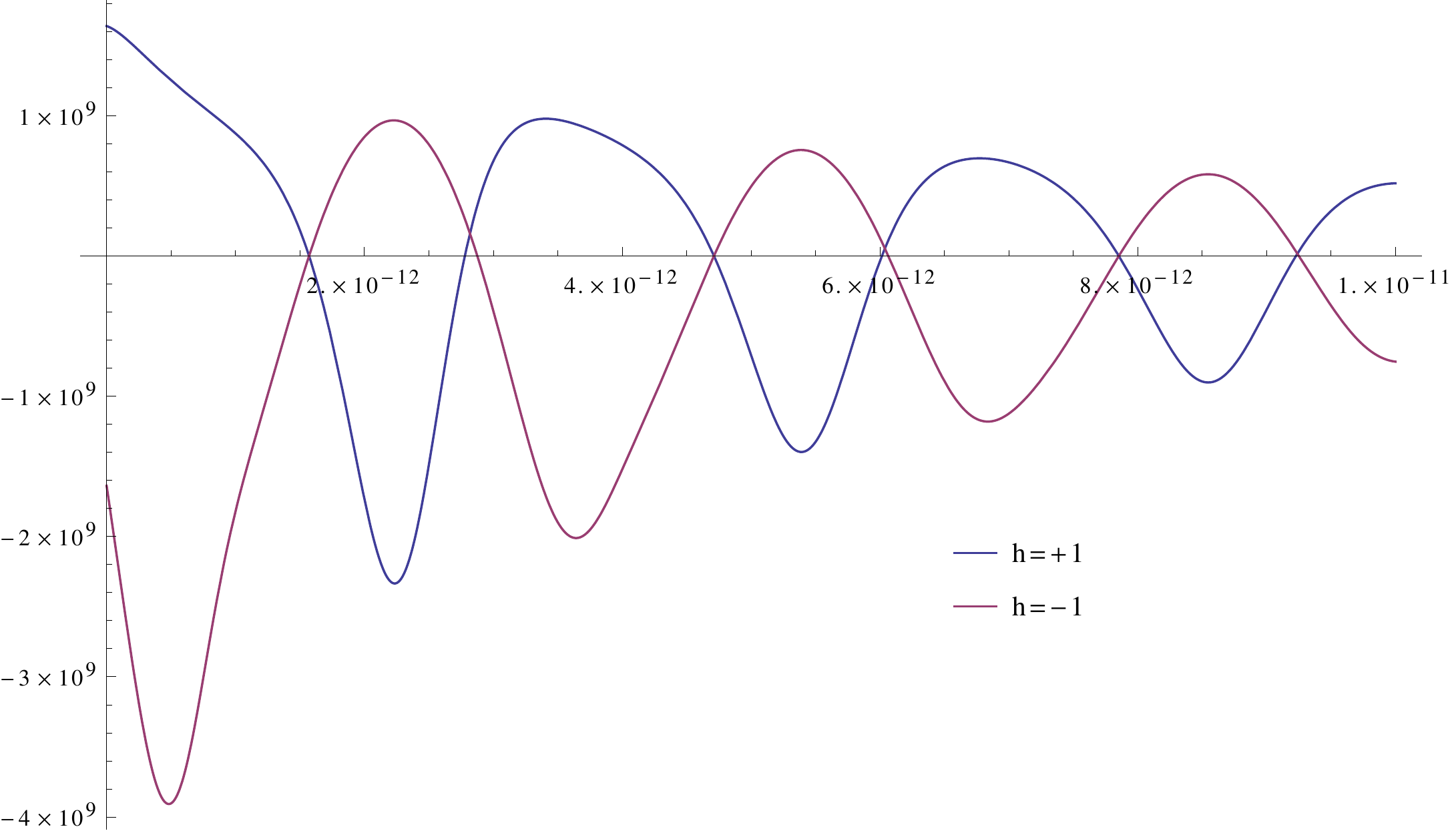}
\caption{$C(t)$ for $v_0 = M = \Omega = 10^{12}$~GeV and $\Gamma_H = 10^{11}$~GeV, which governs the differential equations for $\bar{\beta}$.  Units are $\mathrm{GeV}^{-1}$ for time and $\mathrm{GeV}$ for $C(t)$.}
\label{fig:covera_model}
\end{figure}

As a numerical example, we used $v_0 = M = \Omega = 10^{12}$~GeV and $\Gamma_H = 10^{11}$~GeV, which gives $\mu_\mathrm{max} \approx 4.6 \cdot 10^{11}$~GeV. A plot of $C(t)$ is shown in Fig.~\ref{fig:covera_model}; the above-mentioned spikes are small for these parameters.

$\bar{\beta}$ asymptotically approaches 0.0008 (for $h=+1$) and 0.006 (for $h=-1$).  We have verified that $\bar{\beta} < .1$ at all times, so that our approximation in equation~\eqref{eq:approx_beta_DE} is reasonable.  The resulting asymmetry is
\begin{align}
\eta_L \approx \dfrac{10^{28} \; \mathrm{GeV}^3}{a(t)^3 T(t)^3}.
\end{align}
We must ensure that the Higgs energy density does not dominate the energy density of the universe, causing additional inflationary expansion, which requires $\Lambda_I \gtrsim 10^{12} \; \mathrm{GeV}$.  For the minimum value, the inflationary Hubble parameter is then $H_I = 2 \cdot 10^5 \; \mathrm{GeV}$.  If we take $\Gamma_I = 10^5 \; \mathrm{GeV}$, then at late times $a(t) T(t) \rightarrow 5 \cdot 10^{11} \; \mathrm{GeV}$.  The resulting lepton asymmetry $\eta_L \approx 10^{-7}$.  

We note that as $m \sim \Omega \gg H_I$, a realistic implementation of these parameters would likely have the Higgs VEV trapped in a false vacuum.  The relaxation of the Higgs field would then commence after the start of coherent oscillations, or $t_S > t_{\mathrm{end} \; \mathrm{of} \; \mathrm{inflation}}$.  The asymmetry is then enhanced by the factor $a(t_S)^3 \slash a(t_{\mathrm{end} \; \mathrm{of} \; \mathrm{inflation}})^3 > 1$.

This asymmetry will be diluted by a factor of 30 due to further entropy production, and it will be distributed between baryons and leptons by electroweak sphalerons.  Therefore final baryonic asymmetry is about one or two orders of magnitude smaller.  

In summary, we have shown an explicit numerical example in which the asymmetry generated through neutrino production is more than sufficient to explain the cosmological baryon abundance.  We have also seen that production of a large asymmetry requires significant damping of the oscillations of the Higgs VEV, and also favors a low inflationary scale.

\section{Conclusions}
\label{sec:Conclusions}

In this work, we have explored particle production during an epoch of post-inflationary Higgs relaxation, with a particular emphasis on the production of a lepton asymmetry, which can be converted into the observed baryonic asymmetry.  Unlike in previously considered models~\cite{Kusenko:2014lra,Yang:2015ida}, the asymmetry considered here is produced via the evolution Higgs condensate directly, and does not involve interactions in the plasma produced by inflaton decay.  Therefore, these models do not require a fast reheating, and in fact, we have shown a low reheating scale is desirable.

In particular, we have introduced a specific $\mathcal{O}_6$ operator which involves only Standard Model fields (although extensions of the Standard Model may be necessary to produce this operator).  This operator produces an effective chemical potential for lepton number.

We have solved the equations of motion exactly, including both this operator and a time-dependent Majorana mass.  We then used a Bogoliubov transformation to relate the time-dependent creation and annihilation operators to the corresponding operators fixed at the time when the Higgs relaxation began.  The resulting Bogoliubov coefficients describe the rate of neutrino production during Higgs relaxation.  From this, we calculated the resulting lepton asymmetry.

After completing this formal analysis, we performed a numerical analysis, using a simplified model for the Higgs condensate evolution.  This emphasized the importance of rapid condensate decay, which suppresses washout due to the oscillating sign of the effective chemical potential, and also the low reheat scale.  We developed an approximation scheme that smooths out the sharp peaks that occur when $|\boldsymbol p| \approx \mu(t)$.  We finally illustrated a choice of parameters for which the resulting asymmetry is comparable to the observed value.

Our scenario differs significantly from other scenarios of leptogenesis.  In particular, the asymmetry can be generated for reheat temperatures well below the right-handed neutrino masses.  This paves the way for a supersymmetric generalization of the model in which the problem of gravitino overproduction does not arise.  Furthermore, the final asymmetry is not tied to the parameters of the neutrino mass matrix as in thermal leptogenesis, and a successful leptogenesis is possible even for the neutrino masses above 0.2 eV, in which case  thermal leptogenesis is stymied by excessive washout \cite{Buchmuller:2005eh}.

\section*{Acknowledgements}

The authors wish to thank M.\ Ibe, K.\ Schmitz, F.\ Takahashi, M.\ Voloshin, and T.\ T.\ Yanagida for helpful discussions.  A.K.\ was supported by the U.S. Department of Energy Grant No. 
DE-SC0009937 and by the World Premier International Research
Center Initiative (WPI Initiative), MEXT, Japan.  The work of M.P.\ is partially supported from the DOE grant DE-SC0011842 at the University of Minnesota.

\appendix

\section{The Origin of the $\mathcal{O}_6$ Operator}
\label{ap:Generating_O_6_operator}

In this appendix, we discuss methods of generating the $\mathcal{O}_6$ operator
\begin{align}
\mathcal{L}_{\mathcal{O}_6} &= -\dfrac{\Phi^2}{M^2} \partial_\mu j_{B+L}^\mu.
\label{eq:appendix_O6}
\end{align}
In the Standard Model in a flat static spacetime, the ABJ anomaly allows the relation
\begin{align}
\partial_\mu j_{B+L}^\mu = n_g \left( \dfrac{g_2^2}{32 \pi^2} \epsilon^{\mu \nu \alpha \beta} \mathsf{A}^{a}_{\mu \nu} \mathsf{A}^{a}_{ \alpha \beta} - \dfrac{g_1^2}{32 \pi^2} \epsilon^{\mu \nu \alpha \beta} \mathsf{B}_{\mu \nu} \mathsf{B}_{\alpha \beta} \right),
\label{eq:EW_anomaly}
\end{align}
where $\mathsf{A}$ and $\mathsf{B}$ are that $\mathrm{SU}_\mathrm{L}(2)$ and $\mathrm{U}_\mathrm{Y}(1)$ gauge fields, respectively, and $n_g$ is the number of fermion generations.  The substitution of Eq.~\eqref{eq:EW_anomaly} into Eq.~\eqref{eq:appendix_O6} is valid when the decay of electroweak sphalerons is fast, as compared to the Hubble parameter.  Otherwise, the term \ref{eq:appendix_O6} involves the Chern-Simons number density, which is not changed by Higgs relaxation unless the phase of the Higgs VEV evolves.

As to coupling these gauge fields to the Higgs field, we note that an effective term of precisely this form can be generated within the Standard Model, using quark loops and the CP-violating phase of the CKM matrix~\cite{Shaposhnikov:1987tw,Shaposhnikov:1987pf}.  This term is small due to the small Yukawa couplings and small CP-violating phase.  However, such a term can also be generated by heavier states with a different source of CP violation.  The scale in the denominator may be the temperature, due to thermal loops, or the mass scale of new physics~\cite{Shaposhnikov:1987tw,Shaposhnikov:1987pf,Smit:2004kh,Brauner:2012gu}.

The sphaleron transition rate per unit volume at finite temperature, for constant Higgs VEVs, is
\begin{equation}
\Gamma_\mathrm{sp} = k \alpha_W^5 T^4 \exp(-M_W \slash g_W T),
\end{equation}
where the exponential factor accounts for the suppression due to being in the broken phase; it is equivalent to $\exp(-v \slash 2 T)$ where $v$ is the Higgs VEV.  Electroweak sphalerons are in equilibrium when this is greater than $H^4$, where $H$ is the Hubble parameter.  The transition rate in the presence of a quickly evolving Higgs VEV has not been explored, although the rate during the electroweak phase transition from $v = 0$ to $v = 247 \; \mathrm{GeV}$ has been analyzed on the lattice, as a function of $v(T)$~\cite{D'Onofrio:2012jk}.

In section \ref{sec:Operators}, we found that the asymmetry is suppressed by a factor of $(a(t) T(t))^3$, which favors a low inflationary scale.  This generally corresponds to a slow reheating, while Higgs relaxation frequently occurs on a faster time scale.  Therefore, during much of the relaxation period, $v \gtrsim T$ and the sphalerons may not be in thermal equilibrium; the conditions for electroweak sphalerons to be in thermal equilibrium in the presence of a time-dependent background have not been extensively explored.

In each oscillation of the Higgs VEV, there is a brief period as the VEV passes zero during which $v \lesssim T$, during which the above-mentioned suppression is absent.  This also corresponds to the time of maximal particle production, which occurs when $\tilde{v} = a(t) v(t) \approx 0$.  However, the time when the maximal asymmetry is produced is slightly offset from this time, as the effective chemical potential $\tilde{\mu} \propto \tilde{v} \tilde{v}^\prime$ is zero when $\tilde{v} = 0$.  It seems unlikely that the time scale of sphaleron transitions will be less than the relevant time scale during which $v \lesssim T$, even if the time of maximal asymmetry production is within this period.  At the very least, it is difficult to arrange for this to hold.

Therefore we note that, if there is another gauge group which couples chirally to leptons, it will also contribute to the divergence in equation \eqref{eq:EW_anomaly}.  (The chiral coupling is necessary due to Furry's Theorem.)  Provided that interactions between the gauge field configurations and fermions are in thermal equilibrium, we find
\begin{align}
\partial_\mu j^\mu_{B+L} = (\mathrm{EW} \; \mathrm{anomaly}) + \dfrac{n_g \mathcal{C} g^2}{32 \pi^2} \epsilon_{\alpha \beta \mu \nu} \mathsf{F}^{\mu \nu} \mathsf{F}^{\alpha \beta},
\label{eq:divergence}
\end{align}
where $\mathsf{F}$ is the new gauge field and $\mathcal{C}$ is a constant determined by the charges of the leptons and baryons under the new gauge group.  Provided that these gauge bosons acquire masses which are not proportional to the Higgs VEV, it is possible for these to be in thermal equilibrium at the relevant temperatures.  (There may dynamical symmtry breaking in this sector, via a separate Higgs mechanism, or in the case of a $U(1)$ symmetry, via the St\"uckleberg mechanism.)  This equation can be rewritten as,
\begin{align}
\dfrac{n_g \mathcal{C} g^2}{32 \pi^2} \epsilon_{\alpha \beta \mu \nu} \mathsf{F}^{\mu \nu} \mathsf{F}^{\alpha \beta} &= \partial_\mu j^\mu_{B+L-CS},
\end{align}
where $j_{CS}$ is the current associated with the electroweak Chern-Simons charge density.  If the electroweak sphalerons are out of equilibrium, this is conserved, and therefore has no effect on the analysis of sections \ref{sec:Effective_Lagrangian} through \ref{sec:Numerics} (similarly to how the baryonic current has no effect).

Therefore, if the electroweak sphaleron rate is insufficient, we can couple the Higgs boson to a new gauge field combination, $\epsilon_{\alpha \beta \mu \nu} \mathsf{F}^{\mu \nu} \mathsf{F}^{\alpha \beta}$, to generate a term similar to \eqref{eq:appendix_O6}.  As in the electroweak case, the coupling of $\Phi^2$ to $\epsilon_{\alpha \beta \mu \nu} \mathsf{F}^{\mu \nu} \mathsf{F}^{\alpha \beta} $ can be accomplished through either thermal loops or heavy fermions.  In the latter case, it is important that the fermions do not acquire masses through the Standard Model Higgs mechanism; otherwise, the Higgs VEV dependence cancels out.  Such fermions may have soft masses similar to higgsinos and gauginos in supersymmetric models, or if a different Higgs sector is used to give masses to the $\mathsf{F}$ gauge boson, this field may also give masses to the relevant fermions.

The divergence equation \eqref{eq:divergence} holds only in static, flat spacetime; the situation is more complicated in a curved and/or expanding spacetime.  Generically, there may be contributions on the right hand side of the anomaly equation, proportional to the gravitational anomaly~\cite{Dobado:1995ec}.  

If there are $N n_g$ right-handed neutrinos present, then (generalizing the results of~\cite{Dobado:1995ec})
\begin{widetext}
\begin{align}
\dfrac{n_g \mathcal{C} g^2}{32 \pi^2} \epsilon_{\alpha \beta \mu \nu} \mathsf{F}^{\mu \nu} \mathsf{F}^{\alpha \beta} &=  \nabla_\mu j^\mu_{B+L-CS} - \dfrac{n_g}{32 \pi^2} \left(1 - N \right) \left( - \dfrac{\epsilon^{\alpha \beta \gamma \delta}}{24} R_{\mu \nu \alpha \beta} R^{\mu \nu}\phantom{}_{\gamma \delta} + \dfrac{\epsilon^{\alpha \beta \gamma \delta}}{48} S_{\beta; \gamma} S_{\delta ; \alpha}  + \dfrac{1}{6} \square S^\alpha_{;\alpha} + \dfrac{1}{96} (S^\alpha S^\nu S_\alpha)_{;\nu}  \right.  \nonumber \\
& \left.- \dfrac{1}{6} \left( R^{\nu \alpha}S_\alpha - \dfrac{1}{2}  R S^\nu \right)_{;\nu} \right),
\end{align}
\end{widetext}
where $S$ describes the torsion of the spacetime, and $R$ is the usual Ricci scalar.  However, if there are the same number of right-handed and left-handed neutrinos then

\begin{align}
\dfrac{n_g \mathcal{C} g^2}{32 \pi^2} \epsilon_{\alpha \beta \mu \nu} \mathsf{F}^{\mu \nu} \mathsf{F}^{\alpha \beta} &= \nabla_\mu j^\mu_{B+L-CS}
\end{align}
We consider only the scenario with $N=1$; that is, there are the same number of right-handed and left-handed neutrinos.

\section{Conformal Higgs Field Equation of Motion}
\label{ap:vhat_eqn_motion}

Although we will use a toy model for our numerical analysis, it is beneficial to find the equation of motion for the comoving VEV $\tilde{v}$.  From the Lagrangian in section \ref{sec:Lagrangian},
\begin{align}
\tilde{v}^{\prime \prime} - \dfrac{a^{\prime \prime}}{a} \tilde{v} + \dfrac{\partial \tilde{V}}{\partial \tilde{v}} = 0,
\end{align}
where the derivatives signified with a prime are with respect to $\eta$ and  $\tilde{V} = a^4 V$.  We note that this is equivalent to the differential equation for the Higgs VEV $v$
\begin{align}
\dfrac{d^2 v}{dt^2} + 3 H \dfrac{dv}{dt} + \dfrac{\partial V}{\partial v} = 0.
\end{align}
It is necessary to express the potential in terms of comoving fields; as an example, we will do this with the 1-loop Standard Model Higgs potential, including finite temperature corrections.  (However, in our numerical analysis, we will make use of a simpler effective potential for the evolution of the Higgs field.)  The one-loop, zero temperature potential $V$, times $a^4$, can be written
\begin{widetext}
\begin{align}
&\tilde{V}_\phi^\mathrm{1-loop} = \dfrac{a^2}{2} m_\phi^2 \tilde{v}^2 + \dfrac{\lambda}{4} \tilde{v}^4 + \dfrac{1}{(4\pi)^2} \left[ \dfrac{a^4 m_H(v)^4}{4} \left( \ln \left( \dfrac{m_H(v)^2}{S^2} \right) - \dfrac{3}{2} \right) 
+ \dfrac{3 a^4 m_G(v)^4}{4} \left( \ln \left( \dfrac{m_G(v)^2}{S^2} \right) - \dfrac{3}{2} \right) \right. \nonumber \\
&\left. + \dfrac{3 a^4  m_W(v)^4}{2} \left( \ln \left( \dfrac{m_W(v)^2}{S^2} \right) - \dfrac{5}{6} \right)
+ \dfrac{3 a^4  m_Z(v)^4}{4} \left( \ln \left( \dfrac{m_Z(v)^2}{S^2} \right) - \dfrac{5}{6} \right)
- 3 a^4  m_t(v)^4 \left( \ln \left( \dfrac{m_t(v)^2}{S^2} \right) - \dfrac{3}{2} \right)
\right],
\end{align}
\end{widetext}
where $S$ is the renormalization scale and the physical masses for the Higgs boson, Goldstone mode, $W$ bosons, $Z$ boson, and top masses are
\begin{alignat}{3}
m_W^2 &= \dfrac{g^{2} v^2}{4}, &\quad m_Z^2 &= \dfrac{(g^2 + g^{\prime \, 2}) v^2}{4}, \quad &
m_t &= \dfrac{y_t v}{\sqrt{2}}, \nonumber \\
m_H^2 &= m_\phi^2 + 3 \lambda  v^2, &\quad m_G^2 &= m_\phi^2 + \lambda v^2.
\end{alignat}

It is convenient to define a comoving renormalization scale, $\tilde{S} = a S$, along with comoving masses
\begin{align}
\tilde{m}_W^2 &= a^2 \dfrac{g^2 v^2}{4} = \dfrac{g^2 \tilde{v}^2}{4} \nonumber \\
\tilde{m}_Z^2 &= \dfrac{(g^2 + g^{\prime 2})a^2 v^2}{4} = \dfrac{(g^2 + g^{\prime 2})^2 \tilde{v}^2}{4} \nonumber \\
\tilde{m}_t &= a \dfrac{y_t v}{\sqrt{2}} = \dfrac{y_t \tilde{v}}{\sqrt{2}} \nonumber \\
\tilde{m}_H^2 &= a^2 (m_\phi^2 + 2 \lambda v^2) = \tilde{m}_\phi^2 + 2 \lambda \tilde{v}^2 \nonumber \\
\tilde{m}_G^2 &= a^2 (m_\phi^2 + \lambda v^2) = \tilde{m}_\phi^2 + \lambda \tilde{v}^2.
\end{align}
These have the same functional dependence on $\tilde{v}$ as the regular masses have on $v$.  Then the one-loop potential can be written:
\begin{widetext}
\begin{align}
\tilde{V}_\phi^\mathrm{1-loop} &= \dfrac{1}{2} \tilde{m}_\phi^2 \tilde{v}^2 + \dfrac{\lambda}{4} \tilde{v}^4 + \dfrac{1}{(4\pi)^2} \left[ \dfrac{\tilde{m}_H(\tilde{v})^4}{4} \left( \ln \left( \dfrac{\tilde{m}_H(\tilde{v})^2}{\tilde{S}^2} \right) - \dfrac{3}{2} \right) 
+ \dfrac{3 \tilde{m}_G(\tilde{v})^4}{4} \left( \ln \left( \dfrac{\tilde{m}_G(\tilde{v})^2}{\tilde{S}^2} \right) - \dfrac{3}{2} \right) \right. \nonumber \\
&\left. + \dfrac{3 \tilde{m}_W(\tilde{v})^4}{2} \left( \ln \left( \dfrac{\tilde{m}_W(\tilde{v})^2}{\tilde{S}^2} \right) - \dfrac{5}{6} \right)
+ \dfrac{3 \tilde{m}_Z(\tilde{v})^4}{4} \left( \ln \left( \dfrac{\tilde{m}_Z(\tilde{v})^2}{\tilde{S}^2} \right) - \dfrac{5}{6} \right)
- 3 \tilde{m}_t(v)^4 \left( \ln \left( \dfrac{\tilde{m}_t(\tilde{v})^2}{\tilde{S}^2} \right) - \dfrac{3}{2} \right)
\right].
\end{align}
We note that care must be used in evaluating the running couplings as functions of the comoving fields.  During reheating finite temperature corrections may also be relevant; in terms of the comoving fields, these are
\begin{align}
\tilde{V}_T(v,T) &= -\dfrac{a^2 T^2}{2 \pi^2} \left[ 6 a^2 m_W(v)^2 J_B \left( \dfrac{m_W(v)}{T} \right) + 3 a^2 m_Z(v)^2 J_B \left( \dfrac{m_Z(v)}{T} \right)  + 12 a^2 m_t(v)^2 J_F \left( \dfrac{m_t(v)}{T} \right) \right] \nonumber \\
&= -\dfrac{\tilde{T}^2}{2 \pi^2} \left[ 6 \tilde{m}_W(\tilde{v})^2 J_B \left( \dfrac{\tilde{m}_W(\tilde{v})}{\tilde{T}} \right) + 3 \tilde{m}_Z(\tilde{v})^2 J_B \left( \dfrac{\tilde{m}_Z(\tilde{v})}{\tilde{T}} \right)  + 12 \tilde{m}_t(\tilde{v})^2 J_F \left( \dfrac{\tilde{m}_t(\tilde{v})}{\tilde{T}} \right) \right],
\end{align}
\end{widetext}
where
\begin{align}
J_{B}(y) & =\sum_{n=1}^{\infty}\frac{1}{n^{2}}K_{2}(ny),\label{eq:J_B}\\
J_{F}(y) & =\sum_{n=1}^{\infty}\frac{(-1)^{n+1}}{n^{2}}K_{2}(ny).\label{eq:J_F}
\end{align}
and we have defined $\tilde{T} = a T$. Note the first three terms of Eq.\ \eqref{eq:J_B} and \eqref{eq:J_F} are fairly good approximation.

The relevant potential for the comoving VEV is then:
\begin{align}
\tilde{V}(\tilde{v},\tilde{T}) = \tilde{V}_\phi^\mathrm{1-loop}(\tilde{v}) + \tilde{V}_T(\tilde{v},\tilde{T}).
\end{align}
We note that as in~\cite{Kusenko:2014lra,Yang:2015ida}, it may be necessary to add further higher dimensional terms to the potential to produce a quasistable vacuum at large VEVs and/or to suppress isocurvature perturbations due to variations in baryon density.  Additionally, dissipation effects may be relevant, and can also influence the production of a baryon asymmetry~\cite{Bartrum:2014fla}.

\section{Two-Component Spinor Conventions}
\label{ap:Conventions}

In the chiral basis, the Dirac $\gamma$ matrices are
\begin{align}
\gamma^0 = \begin{pmatrix}
0 & 1 \\ 1 & 0
\end{pmatrix}, \qquad \gamma^i = \begin{pmatrix}
0 & \sigma^i \\ - \sigma^i & 0
\end{pmatrix},
\end{align}
and the projection operators are
\begin{align}
P_R &= \begin{pmatrix}
0 & 0 \\ 0 & 1
\end{pmatrix}, \qquad P_L = \begin{pmatrix}
1 & 0 \\ 0 & 0 
\end{pmatrix}.
\end{align}
I note that
\begin{equation}
\gamma^0 P_L = \begin{pmatrix}
0 & 0 \\ 1 & 0
\end{pmatrix}, \qquad \gamma^0 P_R = \begin{pmatrix}
0 & 1 \\ 0 & 0 
\end{pmatrix},
\end{equation}
and  the complex conjugation operator is given by
\begin{align}
\mathcal{C} &= i \gamma^2 \gamma^0 = i \begin{pmatrix} 0 & \sigma_2 \\
- \sigma_2 & 0 \end{pmatrix} \begin{pmatrix} 0 & 1 \\ 1 & 0 \end{pmatrix}
= i \begin{pmatrix} \sigma_2 & 0 \\ 0 & - \sigma_2 \end{pmatrix} .
\end{align}

We also use the four-vector of Pauli matrices,
\begin{align}
\sigma^\mu = (1, \sigma^i) \qquad \bar{\sigma}^\mu = (1, -\sigma^i)
\end{align}

\section{Diagonalizing the Hamiltonian (Constant Mass and Chemical Potential)}
\label{ap:constant_mass_case}

In this appendix, we present the important steps leading from equation \eqref{eq:H_field} to \eqref{eq:H_a_operators}, for the interested reader.  The two terms in \eqref{eq:H_field} can be written as
\begin{widetext}
\begin{align}
\int d^3 x \, \hat{\nu}_L^\dagger \partial_0 \hat{\nu}_L  
&= \int \dfrac{d^3\tilde{k}}{(2\pi)^{3}} \sum_{h,\bar{h}} \left[[u(h,\tilde{k})^* \partial_0 u(\bar{h},\tilde{k})] a^{(h)\dagger}_{\tilde{k}} a^{(\bar{h})}_{\tilde{k}} \chi^{(h)\dagger}_{\tilde{\boldsymbol k}} \chi^{(\bar{h})}_{\tilde{\boldsymbol k}} 
 - [u(h,\tilde{k})^* \partial_0 v(\bar{h},\tilde{k}_D)^*] a^{(h)\dagger}_{\tilde{k}} a^{(\bar{h})\dagger}_{\tilde{k}_D} \chi^{(h)\dagger}_{\tilde{\boldsymbol k}} \chi^{(-\bar{h})}_{-\tilde{\boldsymbol k}} \right. \nonumber \\
& \qquad \left. - [v(h,\tilde{k}) \partial_0 u(\bar{h},\tilde{k}_D)] a^{(h)}_{\tilde{k}} a^{(\bar{h})}_{\tilde{k}_D} \chi^{(-h)\dagger}_{\tilde{\boldsymbol k}} \chi^{(\bar{h})}_{-\tilde{\boldsymbol k}} 
 + [v(h,\tilde{k}) \partial_0 v(\bar{h},\tilde{k})^*] a^{(h)}_{\tilde{k}} a^{(\bar{h})\dagger}_{\tilde{k}} \chi^{(-h)\dagger}_{\tilde{\boldsymbol k}} \chi^{(-\bar{h})}_{\tilde{\boldsymbol k}}\right].
\end{align}
where we use the notation $p_D$ for the four-vector $(E,-\tilde{\boldsymbol k})$.  Similarly,
\begin{align}
\int d^3 x \, (\partial_0 \hat{\nu}_L^\dagger ) \hat{\nu}_L 
&= \int \dfrac{d^3\tilde{k}}{(2\pi)^{3}} \sum_{h,\bar{h}} \left[[ u(\bar{h},\tilde{k}) \partial_0 u(h,\tilde{k})^* ] a^{(h)\dagger}_{\tilde{k}} a^{(\bar{h})}_{\tilde{k}} \chi^{(h)\dagger}_{\tilde{\boldsymbol k}} \chi^{(\bar{h})}_{\tilde{\boldsymbol k}} 
 - [ v(\bar{h},\tilde{k}_D)^* \partial_0 u(h,\tilde{k})^* ] a^{(h)\dagger}_{\tilde{k}} a^{(\bar{h})\dagger}_{\tilde{k}_D} \chi^{(h)\dagger}_{\tilde{\boldsymbol k}} \chi^{(-\bar{h})}_{-\tilde{\boldsymbol k}} \right. \nonumber \\
& \qquad \left. - [u(\bar{h},\tilde{k}_D) \partial_0 v(h,\tilde{k}) ] a^{(h)}_{\tilde{k}} a^{(\bar{h})}_{\tilde{k}_D} \chi^{(-h)\dagger}_{\tilde{\boldsymbol k}} \chi^{(\bar{h})}_{-\tilde{\boldsymbol k}} 
 + [ v(\bar{h},\tilde{k})^* \partial_0 v(h,\tilde{k})] a^{(h)}_{\tilde{k}} a^{(\bar{h})\dagger}_{\tilde{k}} \chi^{(-h)\dagger}_{\tilde{\boldsymbol k}} \chi^{(-\bar{h})}_{\tilde{\boldsymbol k}}\right].
\end{align}
\end{widetext}

%

To evaluate the products of the spinors, we note that
\begin{align}
\chi^{(h)\dagger}_{\tilde{\boldsymbol k}} \chi^{(\bar{h})}_{\tilde{\boldsymbol k}} &= \delta_{h,\bar{h}}, \nonumber \\
\chi^{(-h)\dagger}_{\tilde{\boldsymbol k}} \chi^{(-\bar{h})}_{\tilde{\boldsymbol k}} &= \delta_{-h,-\bar{h}} = \delta_{h,\bar{h}}, \nonumber \\
\chi^{(-h)}(-\tilde{\boldsymbol k}) &= \zeta(\tilde{\boldsymbol k}, h)\chi^{(h)}(\tilde{\boldsymbol k}),
\end{align}
where $\zeta$ is a phase that obeys:
\begin{align}
\zeta(-\tilde{\boldsymbol k},h) &= - \eta(\tilde{\boldsymbol k},h) \nonumber \\
\zeta(-\tilde{\boldsymbol k},-h) &= \eta^*(\tilde{\boldsymbol k},h) \nonumber \\
\zeta(\tilde{\boldsymbol k},-h) &= - \eta^*(\tilde{\boldsymbol k},h).
\end{align}
Additionally using the anticommutation relations, we may write the Hamiltonian as
\begin{widetext}
\begin{align}
&H = \dfrac{i}{2} \int \dfrac{d^3\tilde{k}}{(2\pi)^3} \sum_{h} \left[[u(h,\tilde{k})^* \partial_0 u(h,\tilde{k}) - u(h,\tilde{k}) \partial_0 u(h,\tilde{k})^* - v(h,\tilde{k}) \partial_0 v(h,\tilde{k})^* + v(h,\tilde{k})^*  \partial_0 v(h,\tilde{k})] a^{(h)\dagger}_{\tilde{k}} a^{(\bar{h})}_{\tilde{k}} \right. \nonumber \\
&  \left. - [u(h,\tilde{k})^* \partial_0 v(h,\tilde{k}_D)^* - v(h,\tilde{k}_D)^* \partial_0 u(h,\tilde{k})^*] \zeta(\tilde{\boldsymbol k},h) a^{(h)\dagger}_{\tilde{k}}  a^{(h)\dagger}_{\tilde{k}_D} - [v(h,\tilde{k}) \partial_0 u(h,\tilde{k}_D) - u(h,\tilde{k}_D) \partial_0 v(h,\tilde{k})] \zeta(\tilde{\boldsymbol k},h)^* a^{(h)}_{\tilde{k}_D} a^{(h)}_{\tilde{k}}   \right].
\label{eq:H_us_vs}
\end{align}
\end{widetext}
We note that the $u$'s and $v$'s depend on the momentum only through $|\tilde{\boldsymbol k}|$; therefore, $u(h, \tilde{k}_D) = u(h, \tilde{k})$ and $v(h, \tilde{k}_D) = v(h, \tilde{k})$.  
The first combination is
\begin{align}
&u(h,\tilde{k})^* \partial_0 u(h,\tilde{k}) - u(h,\tilde{k}) \partial_0 u(h,\tilde{k})^* + v(h,\tilde{k})^* \partial_0 v(h,\tilde{k}) \nonumber \\
&\qquad  - v(h,\tilde{k}) \partial_0 v(h,\tilde{k})^* = - 2 i \tilde{\omega} \left[ |\alpha|^2 - |\beta|^2 \right].
\end{align}
while the other two are related by complex conjugation.  One can show
\begin{align}
u(h,\tilde{k}_D)^* \partial_0 v(h,\tilde{k})^* - v(h,\tilde{k})^* \partial_0 u(h,\tilde{k}_D)^*  
&= 2 i \tilde{\omega} h \alpha^* \beta^*.
\end{align}
Together these give Eq.~\eqref{eq:H_a_operators}.

\section{Differential Equations for Bogoliubov Coefficients}
\label{ap:alpha_beta_DEs}

In this appendix, we present the important steps in deriving the differential equations \eqref{eq:alpha_beta_DEs} from the equations of motion.  First, we introduce the notation
\begin{equation}
g_\pm = \sqrt{ 1 \pm f}.
\end{equation}
The equations of motion require
\begin{align}
i \dfrac{du}{d\eta} + h |\tilde{\boldsymbol k}| u + \tilde{\mu}_\mathrm{eff} u &= h \tilde{M}_L v, \nonumber \\
i \dfrac{dv^*}{d\eta} + h |\tilde{\boldsymbol k}| v^* + \tilde{\mu}_\mathrm{eff} v^* &= - h \tilde{M}_L u^*.
\end{align}

Since $\alpha$, $\beta$, and $\tilde{\omega}$ are time-dependent,\footnote{If $F$ is the antiderivative of $\tilde{\omega}$, then $\int_0^\eta \tilde{\omega} d\bar{\eta} = F(\eta) - F(0)$.  Differentiating this with respect to $\eta$ then gives $\tilde{\omega}(\eta)$.}
\begin{widetext}
\begin{align}
\dfrac{du}{d\eta} &= - i \tilde{\omega} \dfrac{\alpha}{\sqrt{2}} g_- e^{-i \int_0^\eta \tilde{\omega} d\bar{\eta}} + i \tilde{\omega} \dfrac{\beta}{\sqrt{2}} g_+ e^{i \int_0^\eta \tilde{\omega} d\bar{\eta}} + \dfrac{du}{d\alpha} \dfrac{d\alpha}{d\eta} + \dfrac{du}{d\beta} \dfrac{d\beta}{d\eta} + \dfrac{\alpha}{\sqrt{2}} \dfrac{dg_-}{d\eta} e^{-i \int_0^\eta \tilde{\omega} d\bar{\eta}} + \dfrac{\beta}{\sqrt{2}} \dfrac{dg_+}{d\eta} e^{i \int_0^\eta \tilde{\omega} d\bar{\eta}} \\
\dfrac{dv}{d\eta} &= - i \tilde{\omega} \dfrac{h\alpha}{\sqrt{2}} g_+ e^{-i \int_0^\eta \tilde{\omega} d\bar{\eta}} - i \tilde{\omega} \dfrac{h \beta}{\sqrt{2}} g_- e^{i \int_0^\eta \tilde{\omega} d\bar{\eta}} + \dfrac{dv}{d\alpha} \dfrac{d\alpha}{d\eta} + \dfrac{dv}{d\beta} \dfrac{d\beta}{d\eta} + \dfrac{h\alpha}{\sqrt{2}} \dfrac{dg_+}{d\eta} e^{-i \int_0^\eta \tilde{\omega} d\bar{\eta}} - \dfrac{h\beta}{\sqrt{2}} \dfrac{dg_-}{d\eta} e^{i \int_0^\eta \tilde{\omega} d\bar{\eta}}
\label{eq:u_v_dt}
\end{align}
However, these functions also satisfy
\begin{align}
- i \tilde{\omega} \dfrac{\alpha}{\sqrt{2}} g_- e^{-i \int_0^\eta \tilde{\omega} d\bar{\eta}} + i \tilde{\omega} \dfrac{\beta}{\sqrt{2}} g_+ e^{i \int_0^\eta \tilde{\omega} d\bar{\eta}} + h |\tilde{\boldsymbol k}| u(h,\tilde{k})  + \tilde{\mu}_\mathrm{eff} u(h,\tilde{k}) &= h \tilde{M}_L v(h,\tilde{k}) \nonumber \\
- i \tilde{\omega} \dfrac{h\alpha}{\sqrt{2}} g_+ e^{-i \int_0^\eta \tilde{\omega} d\bar{\eta}} - i \tilde{\omega} \dfrac{h \beta}{\sqrt{2}} g_- e^{i \int_0^\eta \tilde{\omega} d\bar{\eta}} + h |\tilde{\boldsymbol k}| v(h,\tilde{k})^* + \tilde{\mu}_\mathrm{eff} v(h,\tilde{k})^* &= - h \tilde{M}_L u(h,\tilde{k})^*,
\end{align}
which allows us to simplify the above equations of motion to
\begin{align}
\dfrac{1}{\sqrt{2}} \dfrac{d\alpha}{d\eta} g_- e^{-i \int \int_0^\eta \tilde{\omega} d\bar{\eta}} + \dfrac{1}{\sqrt{2}} \dfrac{d\beta}{d\eta} g_+ e^{i \int_0^\eta \tilde{\omega} d\bar{\eta}} + \dfrac{\alpha}{\sqrt{2}} \dfrac{dg_-}{d\eta} e^{-i \int_0^\eta \tilde{\omega} d\bar{\eta}} + \dfrac{\beta}{\sqrt{2}} \dfrac{dg_+}{d\eta} e^{i \int_0^\eta \tilde{\omega} d\bar{\eta}} &= 0, \nonumber \\
\dfrac{h}{\sqrt{2}} \dfrac{d\alpha}{d\eta} g_+ e^{-i \int_0^\eta \tilde{\omega} d\bar{\eta}} - \dfrac{h 1}{\sqrt{2}} \dfrac{d\beta}{d\eta} g_- e^{i \int_0^\eta \tilde{\omega} d\bar{\eta}}  + \dfrac{h\alpha}{\sqrt{2}} \dfrac{dg_+}{d\eta} e^{-i \int_0^\eta \tilde{\omega} d\bar{\eta}} - \dfrac{h\beta}{\sqrt{2}} \dfrac{dg_-}{d\eta} e^{i \int_0^\eta \tilde{\omega} d\bar{\eta}} &= 0.
\end{align}
We may transform these into equations for the derivatives of $\alpha$ and $\beta$,
\begin{align}
\dfrac{1}{\sqrt{2}} \dfrac{d\alpha}{d\eta} (g_-^2 + g_+^2) e^{-i \int_0^\eta \tilde{\omega} d\bar{\eta}} &= - \dfrac{\alpha}{\sqrt{2}} \left( g_- \dfrac{dg_-}{d\eta} + g_+ \dfrac{dg_+}{d\eta} \right) e^{-i \int_0^\eta \tilde{\omega} d\bar{\eta}} - \dfrac{\beta}{\sqrt{2}} \left( g_- \dfrac{dg_+}{d\eta} - g_+ \dfrac{dg_-}{d\eta} \right) e^{i \int_0^\eta \tilde{\omega} d\bar{\eta}}, \\
\dfrac{1}{\sqrt{2}} \dfrac{d\beta}{d\eta} (g_+^2 + g_-^2) e^{i \int_0^\eta \tilde{\omega} d\bar{\eta}} &= - \dfrac{\alpha}{\sqrt{2}} \left( g_+ \dfrac{dg_-}{d\eta} - g_- \dfrac{dg_+}{d\eta} \right) e^{- i \int_0^\eta \tilde{\omega} d\bar{\eta}} - \dfrac{\beta}{\sqrt{2}}\left( g_+ \dfrac{dg_+}{d\eta} + g_- \dfrac{dg_-}{d\eta}  \right) e^{i \int_0^\eta \tilde{\omega} d\bar{\eta}}.
\label{eq:time_dependent_DEs}
\end{align}
\end{widetext}

We note that 
\begin{align}
g_+^2 + g_-^2 &= 1 + f + 1 - f = 2 ,
\end{align}
and
\begin{align}
g_- \dfrac{dg_-}{d\eta} + g_+ \dfrac{dg_+}{d\eta} 
&= \dfrac{1}{2} \left( -\dfrac{df}{d\eta} + \dfrac{df}{d\eta} \right)= 0.
\end{align}
%

The remaining combination is
\begin{align}
 g_- \dfrac{dg_+}{d\eta} - g_+ \dfrac{dg_-}{d\eta} 
&= \dfrac{1}{\sqrt{ 1 - f^2}} \dfrac{df}{d\eta}.
\end{align}
Therefore, these equations simplify to
\begin{align}
 \dfrac{d\alpha}{d\eta} &= - \dfrac{\beta}{2} \dfrac{1}{\sqrt{ 1 - f^2}} \dfrac{df}{d\eta} e^{2 i \int_0^\eta \tilde{\omega} d\bar{\eta}}, \nonumber \\
\dfrac{d\beta}{d\eta}&= \dfrac{\alpha}{2} \dfrac{1}{\sqrt{ 1 - f^2}} \dfrac{df}{d\eta} e^{- 2 i \int_0^\eta \tilde{\omega} d\bar{\eta}} .
\end{align}
Since $f = (h |\tilde{\boldsymbol k}| + \tilde{\mu}_\mathrm{eff}) \slash \tilde{\omega}$, $\sqrt{ 1 - f^2} = \tilde{M}_L \slash \tilde{\omega}$, using the definition of $\tilde{\omega}$ above.  
Differentiating $f$ gives us
\begin{widetext}
\begin{align}
 \dfrac{d\alpha}{d\eta} &= - \dfrac{\beta}{2} \dfrac{ \tilde{\omega}}{\tilde{M}_L} \left[ \dfrac{1}{\tilde{\omega}} \dfrac{d \tilde{\mu}_\mathrm{eff}}{d\eta} - \dfrac{h|\tilde{\boldsymbol k}| + \tilde{\mu}_\mathrm{eff}}{\tilde{\omega}^3} \left((h|\tilde{\boldsymbol k}| + \tilde{\mu}_\mathrm{eff}) \dfrac{d\tilde{\mu}_\mathrm{eff}}{d\eta} + \tilde{M}_L \dfrac{d\tilde{M}_L}{d\eta} \right) \right] e^{2 i \int_0^\eta \tilde{\omega} d\bar{\eta}}, \nonumber \\
\dfrac{d\beta}{d\eta}&= \dfrac{\alpha}{2} \dfrac{ \tilde{\omega}}{\tilde{M}_L} \left[ \dfrac{1}{\tilde{\omega}} \dfrac{d \tilde{\mu}_\mathrm{eff}}{d\eta} - \dfrac{h|\tilde{\boldsymbol k}| + \tilde{\mu}_\mathrm{eff}}{\tilde{\omega}^3} \left((h|\tilde{\boldsymbol k}| + \tilde{\mu}_\mathrm{eff}) \dfrac{d\tilde{\mu}_\mathrm{eff}}{d\eta} + \tilde{M}_L \dfrac{d\tilde{M}_L}{d\eta} \right) \right] e^{- 2 i \int_0^\eta \tilde{\omega} d\bar{\eta}} .
\end{align}
\end{widetext}
Finally, we can combine the $d\tilde{\mu}_\mathrm{eff} \slash d\eta$ terms, using $\tilde{\omega}^2 - (h|\boldsymbol p| + \tilde{\mu}_\mathrm{eff})^2 = \tilde{M}_L^2$.  This gives Eq.~\eqref{eq:alpha_beta_DEs}, as desired.

Applying equations \eqref{eq:time_dependent_DEs} in equations \eqref{eq:u_v_dt} gives 
\begin{align}
\dfrac{du}{d\eta} &= - i \tilde{\omega} \dfrac{\alpha}{\sqrt{2}} g_- e^{-i \int_0^\eta \tilde{\omega} d\bar{\eta}} + i \tilde{\omega} \dfrac{\beta}{\sqrt{2}} g_+ e^{i \int_0^\eta \tilde{\omega} d\bar{\eta}}  \\
\dfrac{dv}{d\eta} &= - i \tilde{\omega} \dfrac{h\alpha}{\sqrt{2}} g_+ e^{-i \int_0^\eta \tilde{\omega} d\bar{\eta}} - i \tilde{\omega} \dfrac{h \beta}{\sqrt{2}} g_- e^{i \int_0^\eta \tilde{\omega} d\bar{\eta}} ,
\end{align}
which shows that the diagonalization of the Hamiltonian proceeds as in the time-independent case.

\section{Effective Lepton Number Operator}
\label{ap:L_eff}

In this appendix, we derive Eq.~\eqref{eq:Leff_normal_ordered}, starting from Eq.~\eqref{eq:L_eff_def}.  Using the orthonormality of the spinors, anticommutation relations, and the fact that $u(r,\tilde{k}_D) = u(r,\tilde{k})$ and $v(r,\tilde{k}_D) = v(r,\tilde{k})$ because the three-momentum only appears as $|\tilde{\boldsymbol k}|$ inside $u$ and $v$ allows us to write
\begin{align}
\tilde{L}_\mathrm{eff} &=  \int \dfrac{d^3\tilde{k} }{(2\pi)^{3}}  \sum_{r} \left[ (|u|^2 - |v|^2) a_{\tilde{k}}^{(r)\dagger} a_{\tilde{k}}^{(r)} \right. \nonumber \\
& \left. - u^* v^* a^{(r)\dagger}_{\tilde{k}} a^{(r)\dagger}_{\tilde{k}_D} \zeta(\tilde{\boldsymbol k},r) 
+ v u a_{\tilde{k}}^{(r)} a_{\tilde{k}_D}^{(r)} \zeta(\tilde{\boldsymbol k},r)^*\right].
\end{align}

Next we evaluate the products of the $u$'s and $v$'s,
\begin{widetext}
\begin{align}
\tilde{L}_\mathrm{eff} &= \int \dfrac{d^3\tilde{k}}{(2\pi)^{3 }} \sum_{h} \left[ - f \left( |\alpha|^2 - |\beta|^2 \right) a_{\tilde{k}}^{(h)\dagger} a_{\tilde{k}}^{(h)} + \dfrac{\tilde{M}_L}{\tilde{\omega}} \left( \alpha \beta^* e^{-2i \int_0^\eta \tilde{\omega} d\bar{\eta}} + \alpha^* \beta e^{2 i \int_0^\eta \tilde{\omega} d\bar{\eta}} \right) a_{\tilde{k}}^{(h)\dagger} a_{\tilde{k}}^{(h)} \right. \nonumber \\
& \left. - h \left[ \dfrac{\tilde{M}_L}{2\tilde{\omega}} \left(\alpha^2 e^{-2 i \int_0^\eta \tilde{\omega} d\bar{\eta}} - \beta^2 e^{2 i \int_0^\eta \tilde{\omega} d\bar{\eta}} \right) + f \alpha \beta \right]  \zeta(\tilde{\boldsymbol k},h)^* a_{\tilde{k}_D}^{(h)} a_{\tilde{k}}^{(h)}  \right. \nonumber \\
& \left. - h \left[ \dfrac{\tilde{M}_L}{2 \tilde{\omega}} \left(\alpha^{*2} e^{2 i \int_0^\eta \tilde{\omega} d\bar{\eta}} - \beta^{*2} e^{-2 i \int_0^\eta \tilde{\omega} d\bar{\eta}} \right) + f \alpha^* \beta^* \right]  \zeta(\tilde{\boldsymbol k},h) a_{\tilde{k}}^{(h)\dagger} a_{\tilde{k}_D}^{(h)\dagger}  \right]
\end{align}
where $f$ is $(h|\tilde{\boldsymbol k}| - \tilde{\mu}_\mathrm{eff}) \slash \tilde{\omega}$ as above.  Using the transformation equations, we recognize that this is
\begin{align}
\tilde{L}_\mathrm{eff} &= \int \dfrac{d^3\tilde{k}}{(2\pi)^{3 }} \sum_{h} \left[ - f A^{(h)\dagger}_{\tilde{k}} A^{(h)}_{\tilde{k}} \right] + \Delta L_\mathrm{eff},
\label{eq:L_eff_Delta}
\end{align}
where
\begin{align}
&\Delta L_\mathrm{eff} = \int \dfrac{d^3\tilde{k}}{(2\pi a)^{3 }} \sum_{h}  \dfrac{\tilde{M}_L}{\tilde{\omega}} \left[  \left( \alpha \beta^* e^{-2i \int_0^\eta \tilde{\omega} d\bar{\eta}} + \alpha^* \beta e^{2 i \int_0^\eta \tilde{\omega} d\bar{\eta}} \right) a_{\tilde{k}}^{(h)\dagger} a_{\tilde{k}}^{(h)} \right. \nonumber \\
& \left. - h  \dfrac{1}{2} \left(\alpha^2 e^{-2 i \int_0^\eta \tilde{\omega} d\bar{\eta}} - \beta^2 e^{2 i \int_0^\eta \tilde{\omega} d\bar{\eta}} \right)  \zeta(\tilde{\boldsymbol k},h)^* a_{\tilde{k}_D}^{(h)} a_{\tilde{k}}^{(h)}   - h  \dfrac{1}{2} \left(\alpha^{*2} e^{2 i \int_0^\eta \tilde{\omega} d\bar{\eta}} - \beta^{*2} e^{-2 i \int_0^\eta \tilde{\omega} d\bar{\eta}} \right) \zeta(\tilde{\boldsymbol k},h) a_{\tilde{k}}^{(h)\dagger} a_{\tilde{k}_D}^{(h)\dagger}  \right].
\end{align}
\end{widetext}
Inverting the transformation equations gives
\begin{align}
\begin{pmatrix}
a_{\tilde{k}}^{(h)\dagger} \\ a_{\tilde{k}_D}^{(h)}
\end{pmatrix} = 
\begin{pmatrix}
\alpha & -h \beta \zeta^* \\
h \beta^* \zeta & \alpha^*
\end{pmatrix} \begin{pmatrix}
A_{\tilde{k}}^{(h)\dagger} \\ A_{\tilde{k}_D}^{(h)}
\end{pmatrix}
\end{align}
Using this, we find
\begin{align}
&\Delta L_{\mathrm{eff}}= \int \dfrac{d^3\tilde{k}}{(2\pi a)^{3}} \dfrac{\tilde{M}_L}{2 \tilde{\omega}} \sum_{h} \nonumber \\
&  \left[  e^{2i \int_0^\eta \tilde{\omega} d\bar{\eta}} \left( - 2 |\alpha|^2 |\beta|^2 - |\beta|^4 - |\alpha|^4 \right)  \zeta A_{\tilde{k}}^{(h)\dagger} A_{\tilde{k}_D}^{(h)\dagger} \right. \nonumber \\
& \left.  +  e^{-2 i \int_0^\eta \tilde{\omega} d\bar{\eta}} \left( - 2 |\alpha|^2 |\beta|^2 - |\alpha|^4 - |\beta|^4 \right) \zeta^* A_{\tilde{k}_D}^{(h)} A_{\tilde{k}}^{(h)}
 \right]
\end{align}
We note that $|\alpha|^4 + 2 |\alpha|^2 |\beta|^2 + |\beta|^4 = (|\alpha|^2 + |\beta|^2)^2 = 1$ by the normalization condition, so this gives
\begin{align}
\Delta L_{\mathrm{eff}} &= -\int \dfrac{d^3\tilde{k}}{(2\pi a)^{3 }} \dfrac{\tilde{M}_L}{\tilde{\omega}} \sum_{h} \left[  e^{2i \int_0^\eta \tilde{\omega} d\bar{\eta}}  \zeta A_{\tilde{k}}^{(h)\dagger} A_{\tilde{k}_D}^{(h)\dagger} \right. \nonumber \\
& \qquad \left. +  e^{-2 i \int_0^\eta \tilde{\omega} d\bar{\eta}}  \zeta^* A_{\tilde{k}_D}^{(h)} A_{\tilde{k}}^{(h)} \right].
\end{align}
Substituting this into Eq.~\eqref{eq:L_eff_Delta} gives Eq.~\eqref{eq:Leff_normal_ordered}.

\bibliography{Bibliography}

\begin{thebibliography}{40}%
\makeatletter
\providecommand \@ifxundefined [1]{%
 \@ifx{#1\undefined}
}%
\providecommand \@ifnum [1]{%
 \ifnum #1\expandafter \@firstoftwo
 \else \expandafter \@secondoftwo
 \fi
}%
\providecommand \@ifx [1]{%
 \ifx #1\expandafter \@firstoftwo
 \else \expandafter \@secondoftwo
 \fi
}%
\providecommand \natexlab [1]{#1}%
\providecommand \enquote  [1]{``#1''}%
\providecommand \bibnamefont  [1]{#1}%
\providecommand \bibfnamefont [1]{#1}%
\providecommand \citenamefont [1]{#1}%
\providecommand \href@noop [0]{\@secondoftwo}%
\providecommand \href [0]{\begingroup \@sanitize@url \@href}%
\providecommand \@href[1]{\@@startlink{#1}\@@href}%
\providecommand \@@href[1]{\endgroup#1\@@endlink}%
\providecommand \@sanitize@url [0]{\catcode `\\12\catcode `\$12\catcode
  `\&12\catcode `\#12\catcode `\^12\catcode `\_12\catcode `\%12\relax}%
\providecommand \@@startlink[1]{}%
\providecommand \@@endlink[0]{}%
\providecommand \url  [0]{\begingroup\@sanitize@url \@url }%
\providecommand \@url [1]{\endgroup\@href {#1}{\urlprefix }}%
\providecommand \urlprefix  [0]{URL }%
\providecommand \Eprint [0]{\href }%
\providecommand \doibase [0]{http://dx.doi.org/}%
\providecommand \selectlanguage [0]{\@gobble}%
\providecommand \bibinfo  [0]{\@secondoftwo}%
\providecommand \bibfield  [0]{\@secondoftwo}%
\providecommand \translation [1]{[#1]}%
\providecommand \BibitemOpen [0]{}%
\providecommand \bibitemStop [0]{}%
\providecommand \bibitemNoStop [0]{.\EOS\space}%
\providecommand \EOS [0]{\spacefactor3000\relax}%
\providecommand \BibitemShut  [1]{\csname bibitem#1\endcsname}%
\let\auto@bib@innerbib\@empty
\bibitem [{\citenamefont {Bunch}\ and\ \citenamefont
  {Davies}(1978)}]{Bunch:1978yq}%
  \BibitemOpen
  \bibfield  {author} {\bibinfo {author} {\bibfnamefont {T.}~\bibnamefont
  {Bunch}}\ and\ \bibinfo {author} {\bibfnamefont {P.}~\bibnamefont {Davies}},\
  }\href {\doibase 10.1098/rspa.1978.0060} {\bibfield  {journal} {\bibinfo
  {journal} {Proc.Roy.Soc.Lond.}\ }\textbf {\bibinfo {volume} {A360}},\
  \bibinfo {pages} {117} (\bibinfo {year} {1978})}\BibitemShut {NoStop}%
\bibitem [{\citenamefont {Linde}(1982)}]{Linde:1982uu}%
  \BibitemOpen
  \bibfield  {author} {\bibinfo {author} {\bibfnamefont {A.~D.}\ \bibnamefont
  {Linde}},\ }\href {\doibase 10.1016/0370-2693(82)90293-3} {\bibfield
  {journal} {\bibinfo  {journal} {Phys.Lett.}\ }\textbf {\bibinfo {volume}
  {B116}},\ \bibinfo {pages} {335} (\bibinfo {year} {1982})}\BibitemShut
  {NoStop}%
\bibitem [{\citenamefont {Starobinsky}\ and\ \citenamefont
  {Yokoyama}(1994)}]{Starobinsky:1994bd}%
  \BibitemOpen
  \bibfield  {author} {\bibinfo {author} {\bibfnamefont {A.~A.}\ \bibnamefont
  {Starobinsky}}\ and\ \bibinfo {author} {\bibfnamefont {J.}~\bibnamefont
  {Yokoyama}},\ }\href {\doibase 10.1103/PhysRevD.50.6357} {\bibfield
  {journal} {\bibinfo  {journal} {Phys.Rev.}\ }\textbf {\bibinfo {volume}
  {D50}},\ \bibinfo {pages} {6357} (\bibinfo {year} {1994})},\ \Eprint
  {http://arxiv.org/abs/astro-ph/9407016} {arXiv:astro-ph/9407016 [astro-ph]}
  \BibitemShut {NoStop}%
\bibitem [{\citenamefont {Enqvist}\ \emph {et~al.}(2013)\citenamefont
  {Enqvist}, \citenamefont {Meriniemi},\ and\ \citenamefont
  {Nurmi}}]{Enqvist:2013kaa}%
  \BibitemOpen
  \bibfield  {author} {\bibinfo {author} {\bibfnamefont {K.}~\bibnamefont
  {Enqvist}}, \bibinfo {author} {\bibfnamefont {T.}~\bibnamefont {Meriniemi}},
  \ and\ \bibinfo {author} {\bibfnamefont {S.}~\bibnamefont {Nurmi}},\ }\href
  {\doibase 10.1088/1475-7516/2013/10/057} {\bibfield  {journal} {\bibinfo
  {journal} {JCAP}\ }\textbf {\bibinfo {volume} {1310}},\ \bibinfo {pages}
  {057} (\bibinfo {year} {2013})},\ \Eprint {http://arxiv.org/abs/1306.4511}
  {arXiv:1306.4511 [hep-ph]} \BibitemShut {NoStop}%
\bibitem [{\citenamefont {Kusenko}\ \emph {et~al.}(2015)\citenamefont
  {Kusenko}, \citenamefont {Pearce},\ and\ \citenamefont
  {Yang}}]{Kusenko:2014lra}%
  \BibitemOpen
  \bibfield  {author} {\bibinfo {author} {\bibfnamefont {A.}~\bibnamefont
  {Kusenko}}, \bibinfo {author} {\bibfnamefont {L.}~\bibnamefont {Pearce}}, \
  and\ \bibinfo {author} {\bibfnamefont {L.}~\bibnamefont {Yang}},\ }\href
  {\doibase 10.1103/PhysRevLett.114.061302} {\bibfield  {journal} {\bibinfo
  {journal} {Phys.Rev.Lett.}\ }\textbf {\bibinfo {volume} {114}},\ \bibinfo
  {pages} {061302} (\bibinfo {year} {2015})},\ \Eprint
  {http://arxiv.org/abs/1410.0722} {arXiv:1410.0722 [hep-ph]} \BibitemShut
  {NoStop}%
\bibitem [{\citenamefont {Yang}\ \emph {et~al.}(2015)\citenamefont {Yang},
  \citenamefont {Pearce},\ and\ \citenamefont {Kusenko}}]{Yang:2015ida}%
  \BibitemOpen
  \bibfield  {author} {\bibinfo {author} {\bibfnamefont {L.}~\bibnamefont
  {Yang}}, \bibinfo {author} {\bibfnamefont {L.}~\bibnamefont {Pearce}}, \ and\
  \bibinfo {author} {\bibfnamefont {A.}~\bibnamefont {Kusenko}},\ }\href@noop
  {} {\  (\bibinfo {year} {2015})},\ \Eprint {http://arxiv.org/abs/1505.07912}
  {arXiv:1505.07912 [hep-ph]} \BibitemShut {NoStop}%
\bibitem [{\citenamefont {Cohen}\ and\ \citenamefont
  {Kaplan}(1987)}]{Cohen:1987vi}%
  \BibitemOpen
  \bibfield  {author} {\bibinfo {author} {\bibfnamefont {A.~G.}\ \bibnamefont
  {Cohen}}\ and\ \bibinfo {author} {\bibfnamefont {D.~B.}\ \bibnamefont
  {Kaplan}},\ }\href {\doibase 10.1016/0370-2693(87)91369-4} {\bibfield
  {journal} {\bibinfo  {journal} {Phys.Lett.}\ }\textbf {\bibinfo {volume}
  {B199}},\ \bibinfo {pages} {251} (\bibinfo {year} {1987})}\BibitemShut
  {NoStop}%
\bibitem [{\citenamefont {Dine}\ \emph {et~al.}(1991)\citenamefont {Dine},
  \citenamefont {Huet}, \citenamefont {Singleton},\ and\ \citenamefont
  {Susskind}}]{Dine:1990fj}%
  \BibitemOpen
  \bibfield  {author} {\bibinfo {author} {\bibfnamefont {M.}~\bibnamefont
  {Dine}}, \bibinfo {author} {\bibfnamefont {P.}~\bibnamefont {Huet}}, \bibinfo
  {author} {\bibfnamefont {J.}~\bibnamefont {Singleton}, \bibfnamefont
  {Robert~L.}}, \ and\ \bibinfo {author} {\bibfnamefont {L.}~\bibnamefont
  {Susskind}},\ }\href {\doibase 10.1016/0370-2693(91)91905-B} {\bibfield
  {journal} {\bibinfo  {journal} {Phys.Lett.}\ }\textbf {\bibinfo {volume}
  {B257}},\ \bibinfo {pages} {351} (\bibinfo {year} {1991})}\BibitemShut
  {NoStop}%
\bibitem [{\citenamefont {Ibe}\ and\ \citenamefont
  {Kaneta}(2015)}]{Ibe:2015nfa}%
  \BibitemOpen
  \bibfield  {author} {\bibinfo {author} {\bibfnamefont {M.}~\bibnamefont
  {Ibe}}\ and\ \bibinfo {author} {\bibfnamefont {K.}~\bibnamefont {Kaneta}},\
  }\href@noop {} {\  (\bibinfo {year} {2015})},\ \Eprint
  {http://arxiv.org/abs/1504.04125} {arXiv:1504.04125 [hep-ph]} \BibitemShut
  {NoStop}%
\bibitem [{\citenamefont {Kusenko}\ \emph {et~al.}(2014)\citenamefont
  {Kusenko}, \citenamefont {Schmitz},\ and\ \citenamefont
  {Yanagida}}]{Kusenko:2014uta}%
  \BibitemOpen
  \bibfield  {author} {\bibinfo {author} {\bibfnamefont {A.}~\bibnamefont
  {Kusenko}}, \bibinfo {author} {\bibfnamefont {K.}~\bibnamefont {Schmitz}}, \
  and\ \bibinfo {author} {\bibfnamefont {T.~T.}\ \bibnamefont {Yanagida}},\
  }\href@noop {} {\  (\bibinfo {year} {2014})},\ \Eprint
  {http://arxiv.org/abs/1412.2043} {arXiv:1412.2043 [hep-ph]} \BibitemShut
  {NoStop}%
\bibitem [{\citenamefont {Birrell}\ and\ \citenamefont
  {Davies}(1982)}]{Birrell:1982ix}%
  \BibitemOpen
  \bibfield  {author} {\bibinfo {author} {\bibfnamefont {N.}~\bibnamefont
  {Birrell}}\ and\ \bibinfo {author} {\bibfnamefont {P.}~\bibnamefont
  {Davies}},\ }\href@noop {} {\bibfield  {journal} {\bibinfo  {journal}
  {Cambridge Monogr.Math.Phys.}\ } (\bibinfo {year} {1982})}\BibitemShut
  {NoStop}%
\bibitem [{\citenamefont {Traschen}\ and\ \citenamefont
  {Brandenberger}(1990)}]{Traschen:1990sw}%
  \BibitemOpen
  \bibfield  {author} {\bibinfo {author} {\bibfnamefont {J.~H.}\ \bibnamefont
  {Traschen}}\ and\ \bibinfo {author} {\bibfnamefont {R.~H.}\ \bibnamefont
  {Brandenberger}},\ }\href {\doibase 10.1103/PhysRevD.42.2491} {\bibfield
  {journal} {\bibinfo  {journal} {Phys.Rev.}\ }\textbf {\bibinfo {volume}
  {D42}},\ \bibinfo {pages} {2491} (\bibinfo {year} {1990})}\BibitemShut
  {NoStop}%
\bibitem [{\citenamefont {Dolgov}\ and\ \citenamefont
  {Kirilova}(1990)}]{Dolgov:1989us}%
  \BibitemOpen
  \bibfield  {author} {\bibinfo {author} {\bibfnamefont {A.}~\bibnamefont
  {Dolgov}}\ and\ \bibinfo {author} {\bibfnamefont {D.}~\bibnamefont
  {Kirilova}},\ }\href@noop {} {\bibfield  {journal} {\bibinfo  {journal}
  {Sov.J.Nucl.Phys.}\ }\textbf {\bibinfo {volume} {51}},\ \bibinfo {pages}
  {172} (\bibinfo {year} {1990})}\BibitemShut {NoStop}%
\bibitem [{\citenamefont {Greene}\ and\ \citenamefont
  {Kofman}(2000)}]{Greene:2000ew}%
  \BibitemOpen
  \bibfield  {author} {\bibinfo {author} {\bibfnamefont {P.~B.}\ \bibnamefont
  {Greene}}\ and\ \bibinfo {author} {\bibfnamefont {L.}~\bibnamefont
  {Kofman}},\ }\href {\doibase 10.1103/PhysRevD.62.123516} {\bibfield
  {journal} {\bibinfo  {journal} {Phys.Rev.}\ }\textbf {\bibinfo {volume}
  {D62}},\ \bibinfo {pages} {123516} (\bibinfo {year} {2000})},\ \Eprint
  {http://arxiv.org/abs/hep-ph/0003018} {arXiv:hep-ph/0003018 [hep-ph]}
  \BibitemShut {NoStop}%
\bibitem [{\citenamefont {Giudice}\ \emph {et~al.}(1999)\citenamefont
  {Giudice}, \citenamefont {Peloso}, \citenamefont {Riotto},\ and\
  \citenamefont {Tkachev}}]{Giudice:1999fb}%
  \BibitemOpen
  \bibfield  {author} {\bibinfo {author} {\bibfnamefont {G.}~\bibnamefont
  {Giudice}}, \bibinfo {author} {\bibfnamefont {M.}~\bibnamefont {Peloso}},
  \bibinfo {author} {\bibfnamefont {A.}~\bibnamefont {Riotto}}, \ and\ \bibinfo
  {author} {\bibfnamefont {I.}~\bibnamefont {Tkachev}},\ }\href {\doibase
  10.1088/1126-6708/1999/08/014} {\bibfield  {journal} {\bibinfo  {journal}
  {JHEP}\ }\textbf {\bibinfo {volume} {9908}},\ \bibinfo {pages} {014}
  (\bibinfo {year} {1999})},\ \Eprint {http://arxiv.org/abs/hep-ph/9905242}
  {arXiv:hep-ph/9905242 [hep-ph]} \BibitemShut {NoStop}%
\bibitem [{\citenamefont {Kofman}\ \emph {et~al.}(1994)\citenamefont {Kofman},
  \citenamefont {Linde},\ and\ \citenamefont {Starobinsky}}]{Kofman:1994rk}%
  \BibitemOpen
  \bibfield  {author} {\bibinfo {author} {\bibfnamefont {L.}~\bibnamefont
  {Kofman}}, \bibinfo {author} {\bibfnamefont {A.~D.}\ \bibnamefont {Linde}}, \
  and\ \bibinfo {author} {\bibfnamefont {A.~A.}\ \bibnamefont {Starobinsky}},\
  }\href {\doibase 10.1103/PhysRevLett.73.3195} {\bibfield  {journal} {\bibinfo
   {journal} {Phys.Rev.Lett.}\ }\textbf {\bibinfo {volume} {73}},\ \bibinfo
  {pages} {3195} (\bibinfo {year} {1994})},\ \Eprint
  {http://arxiv.org/abs/hep-th/9405187} {arXiv:hep-th/9405187 [hep-th]}
  \BibitemShut {NoStop}%
\bibitem [{\citenamefont {Shtanov}\ \emph {et~al.}(1995)\citenamefont
  {Shtanov}, \citenamefont {Traschen},\ and\ \citenamefont
  {Brandenberger}}]{Shtanov:1994ce}%
  \BibitemOpen
  \bibfield  {author} {\bibinfo {author} {\bibfnamefont {Y.}~\bibnamefont
  {Shtanov}}, \bibinfo {author} {\bibfnamefont {J.~H.}\ \bibnamefont
  {Traschen}}, \ and\ \bibinfo {author} {\bibfnamefont {R.~H.}\ \bibnamefont
  {Brandenberger}},\ }\href {\doibase 10.1103/PhysRevD.51.5438} {\bibfield
  {journal} {\bibinfo  {journal} {Phys.Rev.}\ }\textbf {\bibinfo {volume}
  {D51}},\ \bibinfo {pages} {5438} (\bibinfo {year} {1995})},\ \Eprint
  {http://arxiv.org/abs/hep-ph/9407247} {arXiv:hep-ph/9407247 [hep-ph]}
  \BibitemShut {NoStop}%
\bibitem [{\citenamefont {Kofman}\ \emph {et~al.}(1997)\citenamefont {Kofman},
  \citenamefont {Linde},\ and\ \citenamefont {Starobinsky}}]{Kofman:1997yn}%
  \BibitemOpen
  \bibfield  {author} {\bibinfo {author} {\bibfnamefont {L.}~\bibnamefont
  {Kofman}}, \bibinfo {author} {\bibfnamefont {A.~D.}\ \bibnamefont {Linde}}, \
  and\ \bibinfo {author} {\bibfnamefont {A.~A.}\ \bibnamefont {Starobinsky}},\
  }\href {\doibase 10.1103/PhysRevD.56.3258} {\bibfield  {journal} {\bibinfo
  {journal} {Phys.Rev.}\ }\textbf {\bibinfo {volume} {D56}},\ \bibinfo {pages}
  {3258} (\bibinfo {year} {1997})},\ \Eprint
  {http://arxiv.org/abs/hep-ph/9704452} {arXiv:hep-ph/9704452 [hep-ph]}
  \BibitemShut {NoStop}%
\bibitem [{\citenamefont {{Peebles}}(1987)}]{1987Natur.327..210P}%
  \BibitemOpen
  \bibfield  {author} {\bibinfo {author} {\bibfnamefont {P.~J.~E.}\
  \bibnamefont {{Peebles}}},\ }\href {\doibase 10.1038/327210a0} {\bibfield
  {journal} {\bibinfo  {journal} {\nat}\ }\textbf {\bibinfo {volume} {327}},\
  \bibinfo {pages} {210} (\bibinfo {year} {1987})}\BibitemShut {NoStop}%
\bibitem [{\citenamefont {Enqvist}\ and\ \citenamefont
  {McDonald}(1999)}]{Enqvist:1998pf}%
  \BibitemOpen
  \bibfield  {author} {\bibinfo {author} {\bibfnamefont {K.}~\bibnamefont
  {Enqvist}}\ and\ \bibinfo {author} {\bibfnamefont {J.}~\bibnamefont
  {McDonald}},\ }\href {\doibase 10.1103/PhysRevLett.83.2510} {\bibfield
  {journal} {\bibinfo  {journal} {Phys.Rev.Lett.}\ }\textbf {\bibinfo {volume}
  {83}},\ \bibinfo {pages} {2510} (\bibinfo {year} {1999})},\ \Eprint
  {http://arxiv.org/abs/hep-ph/9811412} {arXiv:hep-ph/9811412 [hep-ph]}
  \BibitemShut {NoStop}%
\bibitem [{\citenamefont {Harigaya}\ \emph {et~al.}(2014)\citenamefont
  {Harigaya}, \citenamefont {Kamada}, \citenamefont {Kawasaki}, \citenamefont
  {Mukaida},\ and\ \citenamefont {Yamada}}]{Harigaya:2014tla}%
  \BibitemOpen
  \bibfield  {author} {\bibinfo {author} {\bibfnamefont {K.}~\bibnamefont
  {Harigaya}}, \bibinfo {author} {\bibfnamefont {A.}~\bibnamefont {Kamada}},
  \bibinfo {author} {\bibfnamefont {M.}~\bibnamefont {Kawasaki}}, \bibinfo
  {author} {\bibfnamefont {K.}~\bibnamefont {Mukaida}}, \ and\ \bibinfo
  {author} {\bibfnamefont {M.}~\bibnamefont {Yamada}},\ }\href {\doibase
  10.1103/PhysRevD.90.043510} {\bibfield  {journal} {\bibinfo  {journal}
  {Phys.Rev.}\ }\textbf {\bibinfo {volume} {D90}},\ \bibinfo {pages} {043510}
  (\bibinfo {year} {2014})},\ \Eprint {http://arxiv.org/abs/1404.3138}
  {arXiv:1404.3138 [hep-ph]} \BibitemShut {NoStop}%
\bibitem [{\citenamefont {Degrassi}\ \emph {et~al.}(2012)\citenamefont
  {Degrassi}, \citenamefont {Di~Vita}, \citenamefont {Elias-Miro},
  \citenamefont {Espinosa}, \citenamefont {Giudice} \emph
  {et~al.}}]{Degrassi:2012ry}%
  \BibitemOpen
  \bibfield  {author} {\bibinfo {author} {\bibfnamefont {G.}~\bibnamefont
  {Degrassi}}, \bibinfo {author} {\bibfnamefont {S.}~\bibnamefont {Di~Vita}},
  \bibinfo {author} {\bibfnamefont {J.}~\bibnamefont {Elias-Miro}}, \bibinfo
  {author} {\bibfnamefont {J.~R.}\ \bibnamefont {Espinosa}}, \bibinfo {author}
  {\bibfnamefont {G.~F.}\ \bibnamefont {Giudice}},  \emph {et~al.},\ }\href
  {\doibase 10.1007/JHEP08(2012)098} {\bibfield  {journal} {\bibinfo  {journal}
  {JHEP}\ }\textbf {\bibinfo {volume} {1208}},\ \bibinfo {pages} {098}
  (\bibinfo {year} {2012})},\ \Eprint {http://arxiv.org/abs/1205.6497}
  {arXiv:1205.6497 [hep-ph]} \BibitemShut {NoStop}%
\bibitem [{\citenamefont {Branchina}\ \emph {et~al.}(2015)\citenamefont
  {Branchina}, \citenamefont {Messina},\ and\ \citenamefont
  {Sher}}]{Branchina:2014rva}%
  \BibitemOpen
  \bibfield  {author} {\bibinfo {author} {\bibfnamefont {V.}~\bibnamefont
  {Branchina}}, \bibinfo {author} {\bibfnamefont {E.}~\bibnamefont {Messina}},
  \ and\ \bibinfo {author} {\bibfnamefont {M.}~\bibnamefont {Sher}},\ }\href
  {\doibase 10.1103/PhysRevD.91.013003} {\bibfield  {journal} {\bibinfo
  {journal} {Phys.Rev.}\ }\textbf {\bibinfo {volume} {D91}},\ \bibinfo {pages}
  {013003} (\bibinfo {year} {2015})},\ \Eprint {http://arxiv.org/abs/1408.5302}
  {arXiv:1408.5302 [hep-ph]} \BibitemShut {NoStop}%
\bibitem [{\citenamefont {Hawking}\ and\ \citenamefont
  {Moss}(1982)}]{Hawking:1981fz}%
  \BibitemOpen
  \bibfield  {author} {\bibinfo {author} {\bibfnamefont {S.}~\bibnamefont
  {Hawking}}\ and\ \bibinfo {author} {\bibfnamefont {I.}~\bibnamefont {Moss}},\
  }\href {\doibase 10.1016/0370-2693(82)90946-7} {\bibfield  {journal}
  {\bibinfo  {journal} {Phys.Lett.}\ }\textbf {\bibinfo {volume} {B110}},\
  \bibinfo {pages} {35} (\bibinfo {year} {1982})}\BibitemShut {NoStop}%
\bibitem [{\citenamefont {Starobinsky}(1982)}]{Starobinsky:1982ee}%
  \BibitemOpen
  \bibfield  {author} {\bibinfo {author} {\bibfnamefont {A.~A.}\ \bibnamefont
  {Starobinsky}},\ }\href {\doibase 10.1016/0370-2693(82)90541-X} {\bibfield
  {journal} {\bibinfo  {journal} {Phys.Lett.}\ }\textbf {\bibinfo {volume}
  {B117}},\ \bibinfo {pages} {175} (\bibinfo {year} {1982})}\BibitemShut
  {NoStop}%
\bibitem [{\citenamefont {Vilenkin}\ and\ \citenamefont
  {Ford}(1982)}]{Vilenkin:1982wt}%
  \BibitemOpen
  \bibfield  {author} {\bibinfo {author} {\bibfnamefont {A.}~\bibnamefont
  {Vilenkin}}\ and\ \bibinfo {author} {\bibfnamefont {L.~H.}\ \bibnamefont
  {Ford}},\ }\href {\doibase 10.1103/PhysRevD.26.1231} {\bibfield  {journal}
  {\bibinfo  {journal} {Phys.Rev.}\ }\textbf {\bibinfo {volume} {D26}},\
  \bibinfo {pages} {1231} (\bibinfo {year} {1982})}\BibitemShut {NoStop}%
\bibitem [{\citenamefont {Mamaev}\ \emph {et~al.}(1976)\citenamefont {Mamaev},
  \citenamefont {Mostepanenko},\ and\ \citenamefont {Frolov}}]{Mamaev:1976tq}%
  \BibitemOpen
  \bibfield  {author} {\bibinfo {author} {\bibfnamefont {S.}~\bibnamefont
  {Mamaev}}, \bibinfo {author} {\bibfnamefont {V.}~\bibnamefont
  {Mostepanenko}}, \ and\ \bibinfo {author} {\bibfnamefont {V.}~\bibnamefont
  {Frolov}},\ }\href@noop {} {\bibfield  {journal} {\bibinfo  {journal}
  {Sov.J.Nucl.Phys.}\ }\textbf {\bibinfo {volume} {23}},\ \bibinfo {pages}
  {592} (\bibinfo {year} {1976})}\BibitemShut {NoStop}%
\bibitem [{\citenamefont {Zeldovich}\ and\ \citenamefont
  {Starobinsky}(1972)}]{Zeldovich:1971mw}%
  \BibitemOpen
  \bibfield  {author} {\bibinfo {author} {\bibfnamefont {Y.}~\bibnamefont
  {Zeldovich}}\ and\ \bibinfo {author} {\bibfnamefont {A.~A.}\ \bibnamefont
  {Starobinsky}},\ }\href@noop {} {\bibfield  {journal} {\bibinfo  {journal}
  {Sov.Phys.JETP}\ }\textbf {\bibinfo {volume} {34}},\ \bibinfo {pages} {1159}
  (\bibinfo {year} {1972})}\BibitemShut {NoStop}%
\bibitem [{\citenamefont {Nilles}\ \emph {et~al.}(2001)\citenamefont {Nilles},
  \citenamefont {Peloso},\ and\ \citenamefont {Sorbo}}]{Nilles:2001fg}%
  \BibitemOpen
  \bibfield  {author} {\bibinfo {author} {\bibfnamefont {H.~P.}\ \bibnamefont
  {Nilles}}, \bibinfo {author} {\bibfnamefont {M.}~\bibnamefont {Peloso}}, \
  and\ \bibinfo {author} {\bibfnamefont {L.}~\bibnamefont {Sorbo}},\ }\href
  {\doibase 10.1088/1126-6708/2001/04/004} {\bibfield  {journal} {\bibinfo
  {journal} {JHEP}\ }\textbf {\bibinfo {volume} {0104}},\ \bibinfo {pages}
  {004} (\bibinfo {year} {2001})},\ \Eprint
  {http://arxiv.org/abs/hep-th/0103202} {arXiv:hep-th/0103202 [hep-th]}
  \BibitemShut {NoStop}%
\bibitem [{\citenamefont {Barnaby}\ \emph {et~al.}(2012)\citenamefont
  {Barnaby}, \citenamefont {Moxon}, \citenamefont {Namba}, \citenamefont
  {Peloso}, \citenamefont {Shiu} \emph {et~al.}}]{Barnaby:2012xt}%
  \BibitemOpen
  \bibfield  {author} {\bibinfo {author} {\bibfnamefont {N.}~\bibnamefont
  {Barnaby}}, \bibinfo {author} {\bibfnamefont {J.}~\bibnamefont {Moxon}},
  \bibinfo {author} {\bibfnamefont {R.}~\bibnamefont {Namba}}, \bibinfo
  {author} {\bibfnamefont {M.}~\bibnamefont {Peloso}}, \bibinfo {author}
  {\bibfnamefont {G.}~\bibnamefont {Shiu}},  \emph {et~al.},\ }\href {\doibase
  10.1103/PhysRevD.86.103508} {\bibfield  {journal} {\bibinfo  {journal}
  {Phys.Rev.}\ }\textbf {\bibinfo {volume} {D86}},\ \bibinfo {pages} {103508}
  (\bibinfo {year} {2012})},\ \Eprint {http://arxiv.org/abs/1206.6117}
  {arXiv:1206.6117 [astro-ph.CO]} \BibitemShut {NoStop}%
\bibitem [{\citenamefont {Figueroa}\ \emph {et~al.}(2015)\citenamefont
  {Figueroa}, \citenamefont {Garcia-Bellido},\ and\ \citenamefont
  {Torrenti}}]{Figueroa:2015rqa}%
  \BibitemOpen
  \bibfield  {author} {\bibinfo {author} {\bibfnamefont {D.~G.}\ \bibnamefont
  {Figueroa}}, \bibinfo {author} {\bibfnamefont {J.}~\bibnamefont
  {Garcia-Bellido}}, \ and\ \bibinfo {author} {\bibfnamefont {F.}~\bibnamefont
  {Torrenti}},\ }\href@noop {} {\  (\bibinfo {year} {2015})},\ \Eprint
  {http://arxiv.org/abs/1504.04600} {arXiv:1504.04600 [astro-ph.CO]}
  \BibitemShut {NoStop}%
\bibitem [{\citenamefont {Frasca}(2009)}]{Frasca:2009yp}%
  \BibitemOpen
  \bibfield  {author} {\bibinfo {author} {\bibfnamefont {M.}~\bibnamefont
  {Frasca}},\ }\href {\doibase 10.1142/S021773230903165X} {\bibfield  {journal}
  {\bibinfo  {journal} {Mod.Phys.Lett.}\ }\textbf {\bibinfo {volume} {A24}},\
  \bibinfo {pages} {2425} (\bibinfo {year} {2009})},\ \Eprint
  {http://arxiv.org/abs/0903.2357} {arXiv:0903.2357 [math-ph]} \BibitemShut
  {NoStop}%
\bibitem [{\citenamefont {Buchmuller}\ \emph {et~al.}(2005)\citenamefont
  {Buchmuller}, \citenamefont {Peccei},\ and\ \citenamefont
  {Yanagida}}]{Buchmuller:2005eh}%
  \BibitemOpen
  \bibfield  {author} {\bibinfo {author} {\bibfnamefont {W.}~\bibnamefont
  {Buchmuller}}, \bibinfo {author} {\bibfnamefont {R.}~\bibnamefont {Peccei}},
  \ and\ \bibinfo {author} {\bibfnamefont {T.}~\bibnamefont {Yanagida}},\
  }\href {\doibase 10.1146/annurev.nucl.55.090704.151558} {\bibfield  {journal}
  {\bibinfo  {journal} {Ann.Rev.Nucl.Part.Sci.}\ }\textbf {\bibinfo {volume}
  {55}},\ \bibinfo {pages} {311} (\bibinfo {year} {2005})},\ \Eprint
  {http://arxiv.org/abs/hep-ph/0502169} {arXiv:hep-ph/0502169 [hep-ph]}
  \BibitemShut {NoStop}%
\bibitem [{\citenamefont {Shaposhnikov}(1987)}]{Shaposhnikov:1987tw}%
  \BibitemOpen
  \bibfield  {author} {\bibinfo {author} {\bibfnamefont {M.}~\bibnamefont
  {Shaposhnikov}},\ }\href {\doibase 10.1016/0550-3213(87)90127-1} {\bibfield
  {journal} {\bibinfo  {journal} {Nucl.Phys.}\ }\textbf {\bibinfo {volume}
  {B287}},\ \bibinfo {pages} {757} (\bibinfo {year} {1987})}\BibitemShut
  {NoStop}%
\bibitem [{\citenamefont {Shaposhnikov}(1988)}]{Shaposhnikov:1987pf}%
  \BibitemOpen
  \bibfield  {author} {\bibinfo {author} {\bibfnamefont {M.}~\bibnamefont
  {Shaposhnikov}},\ }\href {\doibase 10.1016/0550-3213(88)90373-2} {\bibfield
  {journal} {\bibinfo  {journal} {Nucl.Phys.}\ }\textbf {\bibinfo {volume}
  {B299}},\ \bibinfo {pages} {797} (\bibinfo {year} {1988})}\BibitemShut
  {NoStop}%
\bibitem [{\citenamefont {Smit}(2004)}]{Smit:2004kh}%
  \BibitemOpen
  \bibfield  {author} {\bibinfo {author} {\bibfnamefont {J.}~\bibnamefont
  {Smit}},\ }\href {\doibase 10.1088/1126-6708/2004/09/067} {\bibfield
  {journal} {\bibinfo  {journal} {JHEP}\ }\textbf {\bibinfo {volume} {0409}},\
  \bibinfo {pages} {067} (\bibinfo {year} {2004})},\ \Eprint
  {http://arxiv.org/abs/hep-ph/0407161} {arXiv:hep-ph/0407161 [hep-ph]}
  \BibitemShut {NoStop}%
\bibitem [{\citenamefont {Brauner}\ \emph {et~al.}(2012)\citenamefont
  {Brauner}, \citenamefont {Taanila}, \citenamefont {Tranberg},\ and\
  \citenamefont {Vuorinen}}]{Brauner:2012gu}%
  \BibitemOpen
  \bibfield  {author} {\bibinfo {author} {\bibfnamefont {T.}~\bibnamefont
  {Brauner}}, \bibinfo {author} {\bibfnamefont {O.}~\bibnamefont {Taanila}},
  \bibinfo {author} {\bibfnamefont {A.}~\bibnamefont {Tranberg}}, \ and\
  \bibinfo {author} {\bibfnamefont {A.}~\bibnamefont {Vuorinen}},\ }\href
  {\doibase 10.1007/JHEP11(2012)076} {\bibfield  {journal} {\bibinfo  {journal}
  {JHEP}\ }\textbf {\bibinfo {volume} {1211}},\ \bibinfo {pages} {076}
  (\bibinfo {year} {2012})},\ \Eprint {http://arxiv.org/abs/1208.5609}
  {arXiv:1208.5609 [hep-ph]} \BibitemShut {NoStop}%
\bibitem [{\citenamefont {D'Onofrio}\ \emph {et~al.}(2012)\citenamefont
  {D'Onofrio}, \citenamefont {Rummukainen},\ and\ \citenamefont
  {Tranberg}}]{D'Onofrio:2012jk}%
  \BibitemOpen
  \bibfield  {author} {\bibinfo {author} {\bibfnamefont {M.}~\bibnamefont
  {D'Onofrio}}, \bibinfo {author} {\bibfnamefont {K.}~\bibnamefont
  {Rummukainen}}, \ and\ \bibinfo {author} {\bibfnamefont {A.}~\bibnamefont
  {Tranberg}},\ }\href {\doibase 10.1007/JHEP08(2012)123} {\bibfield  {journal}
  {\bibinfo  {journal} {JHEP}\ }\textbf {\bibinfo {volume} {1208}},\ \bibinfo
  {pages} {123} (\bibinfo {year} {2012})},\ \Eprint
  {http://arxiv.org/abs/1207.0685} {arXiv:1207.0685 [hep-ph]} \BibitemShut
  {NoStop}%
\bibitem [{\citenamefont {Dobado}\ and\ \citenamefont
  {Maroto}(1996)}]{Dobado:1995ec}%
  \BibitemOpen
  \bibfield  {author} {\bibinfo {author} {\bibfnamefont {A.}~\bibnamefont
  {Dobado}}\ and\ \bibinfo {author} {\bibfnamefont {A.~L.}\ \bibnamefont
  {Maroto}},\ }\href {\doibase 10.1103/PhysRevD.54.5185} {\bibfield  {journal}
  {\bibinfo  {journal} {Phys.Rev.}\ }\textbf {\bibinfo {volume} {D54}},\
  \bibinfo {pages} {5185} (\bibinfo {year} {1996})},\ \Eprint
  {http://arxiv.org/abs/hep-ph/9509227} {arXiv:hep-ph/9509227 [hep-ph]}
  \BibitemShut {NoStop}%
\bibitem [{\citenamefont {Bartrum}\ \emph {et~al.}(2015)\citenamefont
  {Bartrum}, \citenamefont {Berera},\ and\ \citenamefont
  {Rosa}}]{Bartrum:2014fla}%
  \BibitemOpen
  \bibfield  {author} {\bibinfo {author} {\bibfnamefont {S.}~\bibnamefont
  {Bartrum}}, \bibinfo {author} {\bibfnamefont {A.}~\bibnamefont {Berera}}, \
  and\ \bibinfo {author} {\bibfnamefont {J.~G.}\ \bibnamefont {Rosa}},\ }\href
  {\doibase 10.1103/PhysRevD.91.083540} {\bibfield  {journal} {\bibinfo
  {journal} {Phys.Rev.}\ }\textbf {\bibinfo {volume} {D91}},\ \bibinfo {pages}
  {083540} (\bibinfo {year} {2015})},\ \Eprint {http://arxiv.org/abs/1412.5489}
  {arXiv:1412.5489 [hep-ph]} \BibitemShut {NoStop}%
\end{thebibliography}%

\end{document}